\RequirePackage[l2tabu, orthodox]{nag}
\RequirePackage{snapshot}

\documentclass[10pt,onecolumn]{extarticle}

\makeatletter
\if@twocolumn
  \usepackage[dvips,letterpaper,top=0.5in, bottom=0.5in, left=0.75in, right=0.5in,includefoot,heightrounded]{geometry}
\else
  \usepackage[dvips,letterpaper,margin=1in,includefoot,heightrounded]{geometry}
\fi

\usepackage{srcltx}

\usepackage[russian,portuges,english]{babel}

\iflanguage{portuges}
    {\newcommand{\keywordname}{Palavras-chaves}}
    {\newcommand{\keywordname}{Keywords}}

\usepackage{multirow}

\usepackage{booktabs}

\usepackage{graphicx}
\usepackage[table]{xcolor}
\usepackage{amssymb}
\usepackage{amsmath}
\usepackage[normalem]{ulem}

\usepackage{color}
\usepackage{float}

\usepackage[labelformat=simple]{subcaption}

\usepackage{amsfonts}

\usepackage{color}

\usepackage{url}
\usepackage{hyperref}
\usepackage{cite}

\usepackage{abstract}

\usepackage{setspace}
\usepackage{flushend}

\usepackage{multicol}

\usepackage{hyperref}

\usepackage{multirow}

\usepackage[short,12hr]{datetime}
       \usepackage{fouriernc}
\makeatletter

\newcommand{\printtitle}{%
\makeatletter
\if@twocolumn

\twocolumn[%
  \maketitle
  \begin{onecolabstract}
    \myabstract
  \end{onecolabstract}
  \begin{center}
    \small
    \textbf{\keywordname}
    \\\medskip
    \mykeywords
  \end{center}
  \bigskip
]
\saythanks
\else
  \maketitle
  \begin{onecolabstract}
    \myabstract
  \end{onecolabstract}
  \begin{center}
    \small
    \textbf{\keywordname}
    \\\medskip
    \mykeywords
  \end{center}
  \bigskip
  \onehalfspacing
\fi
\makeatother
}

\author{%
V. A. Coutinho%
\thanks{%
V.~A.~Coutinho
was with the
Graduate Program in Electrical Engineering
and
the
Signal Processing Group,
Departamento de Estat\'istica,
Universidade Federal de Pernambuco (UFPE), Brazil,
and currently
is with
the
Department of Statistics and Computer Science,
Universidade Federal Rural de Pernambuco, Brazil.
(e-mail: vitor.coutinho@ufrpe.br)
}
\and
R.~J.~Cintra%
\thanks{%
R. J. Cintra was
with the
Department of Statistics, UFPE,
at the time of the research
and currently
is with
Department of Technology,
CAA,
UFPE, Brazil.
(e-mail: rjdsc@ufpe.br)
}
\and
F.~M.~Bayer%
\thanks{%
F.~M.~Bayer
is with the
Departamento de Estat\'istica
and LACESM,
Universidade Federal de Santa Maria,
Brazil
(e-mail: bayer@ufsm.br)
}
}

\title{%
Low-complexity Multidimensional DCT Approximations}

\newcommand{\myabstract}{%
In this paper,
we introduce
low-complexity multidimensional
discrete cosine transform (DCT) approximations.
Three dimensional DCT
(3D~DCT) approximations are
formalized
in terms of
high-order tensor theory.
The formulation is extended to higher dimensions with arbitrary lengths.
Several multiplierless $8\times 8 \times 8$
approximate methods are proposed
and
the computational complexity is discussed
for the general multidimensional
case.
The proposed methods complexity cost was assessed,
presenting considerably
lower arithmetic operations
when compared with
the exact 3D~DCT.
The proposed approximations
were embedded into 3D~DCT-based video coding scheme
and
a modified quantization step was introduced.
The simulation results
showed that the approximate 3D~DCT coding methods
offer almost identical
output visual quality
when
compared with exact 3D~DCT scheme.
The proposed 3D~approximations
were also
employed as a tool for visual tracking.
The approximate 3D~DCT-based proposed system
performs
similarly to the original exact 3D~DCT-based method.
In general,
the suggested methods showed competitive performance at a considerably lower computational cost.
}

\newcommand{\mykeywords}{%
DCT approximation, multidimensional DCT, three dimensional DCT, 3D~DCT,
approximate DCT, image compression, 3D video compression, visual tracking
}

\date{}

\begin{document}

\printtitle

\section{Introduction}

The discrete cosine transform~(DCT)
is an important tool in several practical applications~\cite{Ahmed1974,ahmed1975}.
For highly correlated one-step Markov data,
the DCT
behaves as an approximation for the
Karhunen-Lo\`eve transform~(KLT)~\cite{Clarke1981},
which is the optimum transformation for
data decorrelation~\cite{britanak2007discrete}.
While the KLT kernel is based on the input data statistical behavior, the DCT kernel is not, which facilitates the design of
fast algorithms that depend only on the transform length~\cite{Blahut2010}.
Consequently,
the DCT is adopted in a multitude of data compression applications~\cite{rao1990discrete},
such as
audio coding~\cite{rao1996techniques},
still image compression~\cite{Wallace1992},
and video coding~\cite{bhaskaran1997}.

Depending on the nature of data,
applications require
the computation of the multidimensional DCT~\cite{Gonzalez2001,Cho1991, Boussakta2004,bozinovic2005,danciu2012,xue2015automatic}.
Multidimensional~DCT algorithms usually take advantage of
the well-known DCT
kernel
separability property~\cite{Gonzalez2001,Boussakta2004,Servais1997,song2013local}
and
successively
apply one dimensional DCT~(1D~DCT)
algorithms
for higher dimensional computation~\cite{Blahut2010,zeng2000multdct}.
Such technique is often referred as
the \emph{row-column} method~\cite{Madisetti2009handbook}.
In the context of still image compression,
e.g. JPEG encoding standard~\cite{Wallace1992},
the two-dimensional DCT (2D~DCT) is applied
as the block transformation~\cite{Rao2001}.
In video coding standards,
such as
MPEG~\cite{Wallace1992,mpeg2},
H.261~\cite{h261}
H.263~\cite{h263},
H.264~\cite{h264},
HEVC~\cite{hevc},
the 2D~DCT is considered for spatial decorrelation of each video frame.

In view of such range of applications,
several fast algorithms for the
1D~DCT
have been proposed~\cite{Chen1977,lee1984new,wang1984fast,hou1987fast,Loeffler1989,fw1992}.
Indeed,
the theoretical minimum
for the multiplicative complexity ~\cite{Heideman1988}
was attained
by the Loeffler DCT algorithm~\cite{Loeffler1989}.
Because exact DCT algorithm design is
a mature field of research,
it is unlikely that
new fast algorithms
for the \emph{exact} 1D~DCT could furnish significant improvement in terms of
computational complexity.
In such scenario,
different techniques to further reduce the
1D~DCT computational cost were considered,
such as
the integer DCT~\cite{fong2012,britanak2007discrete,chen2000video,yokotani2006lossless,hnativ2014integer},
the binDCT techinque based on lifting schemes~\cite{Liang2001,tran2000bindct,chen2002multiplierless},
the DCT
approximations~\cite{haweel2001,lengwehasatit2004scalable,cb2011,bc2012,bas2008,
bas2009,bas2013, Potluri2013,cintra2011integer,Cintra2014-sigpro,madanayake2015low},
the pruned DCT algorithms~\cite{Wang1991,Makkaoui2010, Lecuire2012},
as well as
combined approaches~\cite{kouadria2013low,coutinho2015multiplierless,cintra2015energy,KouadriaMechouek2016}.
In fact,
the HEVC coding standard
adopts the integer 2D~DCT
as a key step for decorrelation~\cite{hevc,hevc1}.
In such scenario,
2D~DCT approximations also have been applied successfully,
achieving
competitive performance at
a lower computational cost~\cite{Potluri2013,coutinho2015multiplierless,cintra2015energy}.
DCT approximations are transformations that present low computational cost
while preserving important DCT properties,
such as energy compaction and
decorrelation capability.
Unlike the exact DCT,
approximations
are not subject to
theoretical lower bounds of multiplicative complexity~\cite{Heideman1988}
and
their design is an open field of research.
Several multiplierless DCT approximations
have been proposed since the introduction of the
pioneer
signed DCT~(SDCT)~\cite{haweel2001}.
State-of-the-art
approximations include:
the Lengwehasatit-Ortega DCT approximation~(LODCT)~\cite{lengwehasatit2004scalable},
the Bouguezel-Ahmad-Swamy~(BAS) series~\cite{bas2008,bas2009, bas2010,bas2013},
the rounded DCT~(RDCT)~\cite{cb2011},
the modified rounded DCT~(MRDCT)~\cite{bc2012},
and the improved approximate DCT~(IADCT) in~\cite{Potluri2013}.

Despite of its wide usage in video compression standards,
the 2D~DCT does not
take into account
the correlation between successive video frames.
In general,
video standards address
temporal correlation
by means of motion estimation algorithms~\cite{le1991mpeg},
which present high computational costs~\cite{Chan1997}.
An alternative to avoid such
complex methods
is
the \emph{interframe coding} approach,
which applies block transformation
to three dimensional arrays~\cite{natarajan1977interframe,ohm2004interframe,saponara2012real}.
Consequently,
three dimensional transformations
emerge as the tool of choice.
Three dimensional DCT~(3D~DCT)
based coding exploits both temporal and spatial correlation of pixels,
since energy compaction property is extended to temporal dimension.
In~\cite{mulla2014image,Saponara2012,
sawant2011balanced,bozinovic2005,bozinovic2003scan,lai2002video,Chan1997,chan1997variable, lee1997quantization,Servais1997},
video compression schemes that divides successive frames into ``cubes'' of pixels and applied them to 3D~DCT are proposed.
Chan and Lee proposed a method to generate quantization ``volume'' for 3D~DCT coefficients~\cite{Chan1997},
instead of using usual quantization matrix~\cite{bhaskaran1997}.
Recently,
the ``SoftCast'' architecture for wireless multicast video transmission was proposed~\cite{softcast2010,jakubczak2010softcast,jakubczak2011softcast}.
Such method applies the 3D~DCT, avoiding motion compensation and differential encoding.
In~\cite{li2016spatiotemporal},
the 3D~DCT spatiotemporal decorrelation properties are exploited
for a novel no-reference video quality assessment method.
The 3D~DCT is also considered as feature for liver image segmentation~\cite{danciu2012},
visual tracking~\cite{li2013visualtracking}
and motion analysis~\cite{bozinovic2005}.
Furthermore,
real-time ultrasonic medical and industrial applications
require computationally efficient three dimensional data compression~\cite{govindan2016hardware,desmouliers2009adaptive}.
DCT-based designs are an alternative for hardware architectures in such context~\cite{cheng2012mpeg,oruklu20063e,cardoso2004performance}.
In~\cite{Boussakta2004},
a 3D vector-radix decimation-in-frequency~(3D~VR~DIF) algorithm which compute
the 3D~DCT directly is proposed,
presenting fewer multiplications operations
than the row-column method.
In~\cite{jacob2015fpga},
an integer 3D~DCT FPGA implementation for video compression is proposed.
Furthermore,
higher dimensional DCT, e.g. the four dimensional DCT~(4D~DCT),
have found practical applicability~\cite{dai2005fast}
in several contexts
such as
light-field rendering~\cite{magnor2000data,levoy1996light},
lumigraph~\cite{gortler1996lumigraph,zhang2000compression},
and video coding~\cite{sang20136d,anitha2011design, xiangwen20044d}.
Fast algorithms for multidimensional DCT were
proposed in~\cite{zeng2000new,chen2003fast,dai2005fast}.
Despite such wide range of applications,
the design of 3D~DCT approximations
considered as low-complexity tools
represent an unexplored field of research.
The integer 3D~DCT approximation proposed in~\cite{jacob2015fpga}
still requires
a large amount of arithmetic operations,
including multiplication operations.
Moreover,
to the best of our knowledge,
3D~DCT approximations
still
lacks a formal mathematical treatment.

The present work addresses
the derivation of multiplierless 3D~DCT approximations.
High-order tensor theory~\cite{lathauwer1998,lathauwer2000best}
is applied to algebraically formulate the approximate 3D~DCT computation.
The concept is extended to the multidimensional case.
To demonstrate the effectiveness
of the sought 3D~DCT approximations,
we aim at applying them to
two major contexts:
(i)~interframe video coding~\cite{natarajan1977interframe,mulla2014image,Saponara2012,
sawant2011balanced,bozinovic2005,bozinovic2003scan,lai2002video,Chan1997,chan1997variable, lee1997quantization,Servais1997}
and
(ii)~visual tracking~\cite{li2013visualtracking,smeulders2014survey,ross2008incremental}.
In these two distinct real-world
problems,
we provide
quantitative evidence
of the appropriateness of
our approach.

The paper is organized as follows.
In Section~\ref{sec:math_back},
the fundamental mathematical background is presented
and
high-order tensor theory is reviewed.
In Section~\ref{sec:3d_dct_approx},
multidimensional DCT approximations
are addressed.
The approximate computation
for the 3D and muldimensional DCT
is formalized
in Section~\ref{sec_3d_dct_approx_math_def}.
The complexity assessment is discussed
in Section~\ref{sec_complexity_assessment}.
A trade-off analysis is presented in Section~\ref{sec_trade_off}.
Section~\ref{sec:interframe_video_coding}
covers interframe video coding
by means of 3D~DCT approximations.
A method to modify the quantization step
in 3D~DCT based video coding is also proposed.
To further validate
the proposed approximations,
in Section~\ref{sec:tracking},
we assess a 3D~DCT approximation
as a tool for visual tracking.
Conclusions are summarized in Section~\ref{sec:conclusion}.

\section{Mathematical Background}
\label{sec:math_back}

In this section,
we review
the necessary mathematical concepts related
to the DCT and tensor theory.

\subsection{1D and 2D~DCT}

The 1D~DCT maps
an $N$-point discrete signal
$\mathbf{x} =
\begin{bmatrix}
x[0]& x[1] & \cdots & x[N-1]
\end{bmatrix}^\top$
into
the $N$-point signal
$\mathbf{X} =
\begin{bmatrix}
X[0]& X[1]& \cdots& X[N-1]
\end{bmatrix}^\top$,
given
by the following relation~\cite{Ahmed1974}:
\begin{align}
\begin{split}
X[k]
\triangleq
\alpha_{N} [k]
\cdot
\sum_{n=0}^{N-1} x[n]
\cdot
\cos
\left(
\frac{\pi(2n+1)k}{2N}
\right)
,
\\
k = 0,1,\ldots,N - 1,
\end{split}
\label{1d_dct_analit}
\end{align}
where
\begin{align}
\alpha_N[k]
\triangleq
\sqrt{\frac{1}{N}}
\cdot
\begin{cases}
1, & \text{if $k = 0$,} \\
\sqrt{2}, & \text{otherwise}.
\end{cases}
\end{align}
Let $\mathbf{A}$ be a 2D~signal
of size \mbox{$N_1\times N_2$},
whose entries are given by $a[n_1,n_2]$, for $n_i = 1,2,\ldots,N_i-1$, and $i=1,2$.
The
entries of the 2D transform-domain signal~$\mathbf{B}$
are
computed according to the following expression~\cite{Cho1991}:
\begin{align}
\begin{split}
b[k_1,k_2]
\triangleq
&
\alpha_{N_1} [k_1]
\cdot
\alpha_{N_2} [k_2]
\cdot
\sum_{n_1=0}^{N_1-1}
\sum_{n_2=0}^{N_2-1}
a[n_1,n_2]
\\ &
\cdot
\cos \left( \frac{\pi(2n_1+1)k_1}{2N_1} \right)
\cdot
\cos \left( \frac{\pi(2n_2+1)k_2}{2N_2} \right),
\\ &
 k_i = 1,2,\ldots,N_i -1
,
\quad i = 1,2
.
\end{split}
\label{2d_dct_analit}
\end{align}

Both the 1D~DCT and the 2D~DCT
can be expressed by means of matrix products.
For the 1D case,
we have:
\begin{align}
\label{1d_dct_matrix}
\mathbf{X} = \mathbf{C}_{N}\cdot \mathbf{x},
\end{align}
where
$\mathbf{C}_N$ is the DCT matrix whose entries
are expressed by:
\begin{align}
\begin{split}
c_N[k,n]
=
\alpha_N[k] \cdot
\cos \left(
\frac{\pi(2n+1)k}{2N} \right),
\quad
\\
k,n=0,1,\ldots,N-1.
\end{split}
\label{dct_matrix_entries}
\end{align}
For the 2D case,
the input 2D signal can be understood as
a matrix $\mathbf{A}$ of size $N_1\times N_2$ and
its associate transformed signal is furnished by:
\begin{align}
\label{2d_dct_matrix}
\mathbf{B} = \mathbf{C}_{N_1} \cdot \mathbf{A} \cdot \mathbf{C}_{N_2}^\top.
\end{align}

\subsection{3D~DCT and High-order Tensor}
\label{subsec:math_back_tensor}

The 3D~DCT
of a discrete
signal
$\mathcal{T}$
with entries
$t[n_1,n_2,n_3]$,
$n_i = 0,1,\ldots,N_i-1$,
for
$i = 1,2,3,$
is given by
the signal $\mathcal{Y}$,
whose entries are given by~\cite{Boussakta2004,li2013visualtracking}:
\begin{align}
\begin{split}
y[k_1,k_2,k_3]
\triangleq
&
\alpha_{N_1} [k_1]
\cdot
\alpha_{N_2} [k_2]
\cdot
\alpha_{N_3} [k_3]
\\ &
\cdot
\sum_{n_1=0}^{N_1-1}
\sum_{n_2=0}^{N_2-1}
\sum_{n_3=0}^{N_3-1}
t[n_1,n_2,n_3]
\\ &
\cdot
\cos \left( \frac{\pi(2n_1+1)k_1}{2N_1} \right)
\cdot
\cos \left( \frac{\pi(2n_2+1)k_2}{2N_2} \right)
\\&
\cdot
\cos \left( \frac{\pi(2n_3+1)k_3}{2N_3} \right),
\\ &
\quad k_i = 1,2,\ldots,N_i-1,
\quad i=1,2,3.
\end{split}
\label{3d_dct_analit}
\end{align}

Vectors and matrices can be modelled as first- and second-order tensors, respectively~\cite{lathauwer2000best}.
Analogously,
a 3D signal can be understood
as a third-order
tensor~\cite{lathauwer1998, northcott2008multilinear, li2013visualtracking }.
A third-order tensor is simply an array
that requires three indices.
Let
$\mathcal{A} \in \mathbb{F}^{N_1 \times N_2 \times \cdots \times N_R}$
be an $R$th-order tensor
whose entries are given by
$a[n_1, n_2, \ldots, n_R]$,
where
$\mathbb{F}$ can be either
the set of the real or complex numbers
and
$n_i = 0,1,\ldots, N_i-1$,
for
$i=1,2,\ldots, R$.
The $i$-mode product of the tensor
$\mathcal{A}$
by a matrix
$\mathbf{M} \in \mathbb{F}^{H\times N_i}$~\cite[p.~xxxv]{Bernstein2009},
denoted by
$\mathcal{A} \times_i \mathbf{M}$,
is defined as a tensor
$\mathcal{B} \in
\mathbb{F}^{N_1 \times N_2 \times \cdots \times N_{i-1}
\times
H
\times
N_{i+1}
\times \cdots \times N_R}$,
whose entries are expressed by:
\begin{align}
\begin{split}
b[n_1, \ldots, n_{i-1}, h , n_{i+1}, \ldots, n_R]
\triangleq
&
\sum_{n_i = 0}^{N_i - 1}
a[n_1, \ldots, n_i, \ldots, n_R]
\\ &
\cdot m[h,n_i]
,
\end{split}
\label{general_tensor_prod}
\end{align}
where
$m[h,n_i]$
are
the entries of $\mathbf{M}$
and $h = 0,1,\ldots,H-1$.

The $i$-mode product formalism provided by~\eqref{general_tensor_prod}
generalizes vector and matrix products.
In fact,
considering
a
column-vector (first-order tensor)
$\mathbf{v} \in \mathbb{F}^{N_1}$
and
matrices (second-order tensor)
$\mathbf{A} \in \mathbb{F}^{N_1 \times N_2}$,
$\mathbf{M}_1 \in \mathbb{F}^{L \times N_1}$, and
$\mathbf{M}_2 \in \mathbb{F}^{H \times N_2}$,
the following expressions hold:
\begin{align}
\mathbf{v} \times_1 \mathbf{M}_1 & = \mathbf{M}_1 \cdot \mathbf{v}, \\
\mathbf{A} \times_1 \mathbf{M}_1 & = \mathbf{M}_1 \cdot \mathbf{A}, \\
\mathbf{A} \times_2 \mathbf{M}_2 & = \mathbf{A} \cdot \mathbf{M}_2^\top.
\end{align}
Furthermore,
the $i$-mode product presents the following properties~\cite{lathauwer2000best,li2013visualtracking}:
\begin{align}
\begin{split}
\left( \mathcal{T} \times_i \mathbf{M} \right) \times_j \mathbf{N}
= &
\left( \mathcal{T}  \times_j \mathbf{N} \right) \times_i \mathbf{M}
\\
= &
\mathcal{T} \times_i \mathbf{M} \times_j \mathbf{N},
\quad
i \neq j.
\end{split}
\label{porp_comut}
\end{align}

Taking into account two matrices
$\mathbf{M} \in \mathbb{F}^{N_i \times L}$
and
$\mathbf{G} \in \mathbb{F}^{L \times N_i}$,
and an $R$th-order tensor
$\mathcal{T} \in \mathbb{F}^{N_1 \times N_2 \times \cdots \times N_i \times \cdots \times N_R}$,
it can be shown that~\cite{lathauwer2000best}:
\begin{align}
\mathcal{T} \times_i \left( \mathbf{M} \cdot \mathbf{G} \right)
=
\mathcal{T} \times_i \mathbf{G} \times_i \mathbf{M}.
\label{porp_dist}
\end{align}
In particular,
we have that:
\begin{align}
\mathcal{T} \times_i \mathbf{I}_{N_i}
=
\mathcal{T},
\label{porp_ident}
\end{align}
where
$\mathbf{I}_{N_i}$ is the identity matrix of order $N_i$.

In view of the above,
the
3D~DCT
can be expressed according to
the $i$-mode products of order tensors
by the DCT matrix.
For the 1D~DCT case,
\eqref{1d_dct_matrix}
becomes
$\mathbf{X} = \mathbf{x} \times_1 \mathbf{C}_N$.
Likewise,
the 2D~DCT in~\eqref{2d_dct_matrix} is given by:
$\mathbf{B} = \mathbf{A} \times_1 \mathbf{C}_{N_1} \times_2 \mathbf{C}_{N_2}$.
Accordingly,
let
$\mathcal{T} \in \mathbb{F}^{N_1 \times N_2 \times N_3}$
be
the input discrete signal as a third-order tensor.
The 3D~DCT is the third-order tensor given by~\cite{li2013visualtracking}:
\begin{align}
\mathcal{Y} =
\mathcal{T} \times_1
\mathbf{C}_{N_1} \times_2
\mathbf{C}_{N_2} \times_3
\mathbf{C}_{N_3}
.
\end{align}

\subsection{DCT Approximations}
\label{sec_sub:DCT_approximations}

The main goal of the DCT approximations
is to achieve similar mathematical properties
relative to the DCT at a significantly lower computational cost.
In general,
an $N$-point
DCT approximation
$\hat{\mathbf{C}}_N$ is given by
the product of a low-complexity matrix
$\mathbf{T}_N$ and a diagonal matrix $\mathbf{S}_N$.
Orthogonality or quasi-orthogonality properties
are ensured by
$\mathbf{S}_N = \sqrt{\left[\operatorname{diag}\left(\mathbf{T}_N \cdot \mathbf{T}_N^\top \right) \right]^{-1}}$,
where $\operatorname{diag}\left(\cdot\right)$
extracts the diagonal elements
of its matrix arguments
returning a diagonal matrix~\cite{Cintra2014-sigpro}.
Thus,
$\hat{\mathbf{C}}_N = \mathbf{S}_N \cdot \mathbf{T}_N$~\cite{haweel2001,lengwehasatit2004scalable,cb2011,bc2012,bas2008,bas2009,bas2013, Potluri2013}.

Considering the one dimensional case,
an $N$-point input vector $\mathbf{x}$ is transformed into
the
$N$-point output vector $\mathbf{X}$
given
by the following expression:
\begin{align}
\label{equation-S-T}
\begin{split}
\mathbf{X} &=
\hat{\mathbf{C}}_N
\cdot
\mathbf{x} \\
&=
\mathbf{S}_N \cdot \mathbf{T}_N
\cdot
\mathbf{x},
\end{split}
\end{align}
where
all matrices are square of order $N$.
The inverse transformation is computed according to
$\mathbf{x} =
\hat{\mathbf{C}}_N^{-1}
\cdot
\mathbf{X}$,
where the inverse matrix is given by~\cite{Cintra2014-sigpro}:
\begin{align}
\hat{\mathbf{C}}_N^{-1} =
\begin{cases}
\mathbf{T}_N^\top \cdot \mathbf{S}_N, & \text{if }\hat{\mathbf{C}}_N \text{ is orthogonal,}\\
\mathbf{T}_N^{-1} \cdot \mathbf{S}_N^{-1}, & \text{otherwise.}
\end{cases}
\label{inverse_approx_matrix}
\end{align}

For 2D~DCT approximations,
an $N \times N$ input matrix $\mathbf{A}$ is submitted to the following transformation:
\begin{align}
\begin{split}
\mathbf{B} & =
\hat{\mathbf{C}}_N
\cdot
\mathbf{A}
\cdot
\hat{\mathbf{C}}_N^\top
\\
&=
\mathbf{S}_N \cdot \mathbf{T}_N
\cdot
\mathbf{A}
\cdot
\mathbf{T}_N^\top \cdot \mathbf{S}_N^\top
\\
&= \left( \mathbf{s}_N \cdot \mathbf{s}_N^\top \right)
\odot
\left(
\mathbf{T}_N
\cdot
\mathbf{A}
\cdot
\mathbf{T}_N^\top
\right)
,
\end{split}
\end{align}
where
$\mathbf{s}_N$ is an $N$-point
column vector containing the diagonal elements of matrix $\mathbf{S}_N$,
$\mathbf{B}$ is the $N \times N$ transform-domain data,
and
$\odot$~denotes the Hadamard product~\cite{horn2012matrix}.
The term
$(\mathbf{s}_N \cdot \mathbf{s}_N^\top)$
is a matrix populated with multiplicative entries;
whereas
the
operation
$(\mathbf{T} _N
\cdot
\mathbf{A}
\cdot
\mathbf{T}_N^\top)$
is often multiplierless
or
of very low computational cost.

In some contexts,
the
term
$(\mathbf{s}_N \cdot \mathbf{s}_N^\top)$
can be merged into a subsequent operation block.
For instance,
in image/video coding,
the quantization step
can fully absorb the complexity
of
$(\mathbf{s}_N \cdot \mathbf{s}_N^\top)$~\cite{britanak2007discrete}.
In this case,
the term $(\mathbf{s}_N \cdot \mathbf{s}_N^\top)$
does not introduce
any
extra arithmetic operation~\cite{haweel2001,lengwehasatit2004scalable,cb2011,bc2012,bas2008,bas2009,bas2013, Potluri2013,Cintra2014-sigpro}.
Table~\ref{tab_dct_approx}
presents several examples of matrices $\mathbf{T}_N$ and $\mathbf{S}_N$
available in literature
for the popular blocklength $N=8$.

\begin{table*}
\centering
\caption{Several approximate 1D DCT methods for $N=8$}
\label{tab_dct_approx}
\begin{tabular}{l c c}
\toprule
Method & $\mathbf{T}_8$ & $\mathbf{S}_8$ \\
\midrule
SDCT~\cite{haweel2001} &
$\left[\begin{smallmatrix}
 1& 1& 1& 1& 1& 1& 1& 1 \\
 1& 1& 1& 1&-1&-1&-1&-1 \\
 1& 1&-1&-1&-1&-1& 1& 1 \\
 1&-1&-1&-1& 1& 1& 1&-1 \\
 1&-1&-1& 1& 1&-1&-1& 1 \\
 1&-1& 1& 1&-1&-1& 1&-1 \\
 1&-1& 1&-1&-1& 1&-1& 1 \\
 1&-1& 1&-1& 1&-1& 1&-1
\end{smallmatrix}\right]$
 &
$\operatorname{diag}\left(\frac{1}{\sqrt{8}} ,\frac{1}{\sqrt{8}} ,\frac{1}{\sqrt{8}} ,\frac{1}{\sqrt{8}} ,\frac{1}{\sqrt{8}} ,\frac{1}{\sqrt{8}}, \frac{1}{\sqrt{8}},\frac{1}{\sqrt{8}} \right). $
 \\
LODCT~\cite{lengwehasatit2004scalable} &
$\left[\begin{smallmatrix}
1 & 1 & 1 & 1 & 1 & 1 & 1 & 1 \\
1 & 1 & 1 & 0 & 0 &-1 &-1 &-1 \\
1 & \frac{1}{2}  & -\frac{1}{2} & -1 &-1 & -\frac{1}{2} & \frac{1}{2} & 1 \\
1 & 0 & -1 & -1 & 1 & 1 & 0 & -1 \\
1 &-1 &-1 & 1 & 1 &-1 &-1 & 1 \\
1 &-1 & 0 & 1 & -1 & 0 & 1 & -1 \\
\frac{1}{2} & -1 & 1 & -\frac{1}{2} & -\frac{1}{2} & 1 & -1 & \frac{1}{2} \\
0 & -1 & 1 & -1 & 1 & -1 & 1 & 0
\end{smallmatrix}\right]$
 & $\operatorname{diag}\left(\frac{1}{\sqrt{8}} ,\frac{1}{\sqrt{6}} ,\frac{1}{\sqrt{5}} ,\frac{1}{\sqrt{6}} ,\frac{1}{\sqrt{8}} ,\frac{1}{\sqrt{6}}, \frac{1}{\sqrt{5}},\frac{1}{\sqrt{6}} \right)$
   \\
   RDCT~\cite{cb2011}&
$ \left[\begin{smallmatrix}
1 & 1 & 1 & 1 & 1 & 1 & 1 & 1 \\
1 & 1 & 1 & 0 & 0 &-1 &-1 &-1 \\
1 & 0 & 0 &-1 &-1 & 0 & 0 & 1 \\
1 & 0 &-1 &-1 & 1 & 1 & 0 &-1 	\\
1 &-1 &-1 & 1 & 1 &-1 &-1 & 1 \\
1 &-1 & 0 & 1 &-1 & 0 & 1 &-1 \\
0 &-1 & 1 & 0 & 0 & 1 &-1 & 0 \\
0 &-1 & 1 &-1 & 1 &-1 & 1 & 0
\end{smallmatrix}\right]$
 & $\operatorname{diag} \left(\frac{1}{\sqrt{8}}, \frac{1}{\sqrt{6}},\frac{1}{2}, \frac{1}{\sqrt{6}}, \frac{1}{\sqrt{8}}, \frac{1}{\sqrt{6}}, \frac{1}{2}, \frac{1}{\sqrt{6}} \right)$
  \\
MRDCT~\cite{bc2012} &
$ \left[\begin{smallmatrix}
1 & 1 & 1 & 1 & 1 & 1 & 1 & 1 \\
1 & 0 & 0 & 0 & 0 & 0 & 0 &-1 \\
1 & 0 & 0 &-1 &-1 & 0 & 0 & 1 \\
0 & 0 &-1 & 0 & 0 & 1 & 0 & 0 \\
1 &-1 &-1 & 1 & 1 &-1 &-1 & 1 \\
0 &-1 & 0 & 0 & 0 & 0 & 1 & 0 \\
0 &-1 & 1 & 0 & 0 & 1 &-1 & 0 \\
0 & 0 & 0 &-1 & 1 & 0 & 0 & 0
\end{smallmatrix}\right]$
 & $\operatorname{diag} \left(\frac{1}{\sqrt{8}}, \frac{1}{\sqrt{2}},\frac{1}{2}, \frac{1}{\sqrt{2}}, \frac{1}{\sqrt{8}}, \frac{1}{\sqrt{2}}, \frac{1}{2}, \frac{1}{\sqrt{2}} \right)$
\\
BAS-2008~\cite{bas2008}&
$\left[\begin{smallmatrix}
 1& 1& 1& 1& 1& 1& 1& 1 \\
 1& 1& 0& 0& 0& 0&-1&-1 \\
 1& \frac{1}{2}&-\frac{1}{2}&-1&-1&-\frac{1}{2}& \frac{1}{2}& 1 \\
 0& 0&-1& 0& 0& 1& 0& 0 \\
 1&-1&-1& 1& 1&-1&-1& 1 \\
 1&-1& 0& 0& 0& 0& 1&-1 \\
 \frac{1}{2}&-1& 1&-\frac{1}{2}&-\frac{1}{2}& 1&-1& \frac{1}{2} \\
 0& 0& 0&-1& 1& 0& 0& 0
\end{smallmatrix}\right]$
 & $\operatorname{diag} \left( \frac{1}{\sqrt{8}},\frac{1}{2},\frac{1}{\sqrt{5}},\frac{1}{\sqrt{2}},\frac{1}{\sqrt{8}},\frac{1}{2},\frac{1}{\sqrt{5}}, \frac{1}{\sqrt{2}} \right)$
 \\%
BAS-2009~\cite{bas2009} &
$\left[\begin{smallmatrix}
 1& 1& 1& 1& 1& 1& 1& 1 \\
 1& 1& 0& 0& 0& 0&-1&-1 \\
 1& 1&-1&-1&-1&-1& 1& 1 \\
 0& 0&-1& 0& 0& 1& 0& 0 \\
 1&-1&-1& 1& 1&-1&-1& 1 \\
 1&-1& 0& 0& 0& 0& 1&-1 \\
 1&-1& 1&-1&-1& 1&-1& 1 \\
 0& 0& 0&-1& 1& 0& 0& 0
\end{smallmatrix}\right]$
 & $\operatorname{diag} \left( \frac{1}{\sqrt{8}},\frac{1}{2},\frac{1}{\sqrt{8}},\frac{1}{\sqrt{2}},\frac{1}{\sqrt{8}},\frac{1}{2},\frac{1}{\sqrt{8}}, \frac{1}{\sqrt{2}} \right)$
   \\
BAS-2013~\cite{bas2013} &
$ \left[\begin{smallmatrix}
 1& 1& 1& 1& 1& 1& 1& 1 \\
 1& 1& 1& 1&-1&-1&-1&-1 \\
 1& 1&-1&-1&-1&-1& 1& 1 \\
 1& 1&-1&-1& 1& 1&-1&-1 \\
 1&-1&-1& 1& 1&-1&-1& 1 \\
 1&-1&-1& 1&-1& 1& 1&-1 \\
 1&-1& 1&-1&-1& 1&-1& 1 \\
 1&-1& 1&-1& 1&-1& 1&-1
\end{smallmatrix}\right]$
 & $\operatorname{diag} \left(\frac{1}{\sqrt{8}}, \frac{1}{\sqrt{8}},\frac{1}{\sqrt{8}}, \frac{1}{\sqrt{8}}, \frac{1}{\sqrt{8}}, \frac{1}{\sqrt{8}}, \frac{1}{\sqrt{8}}, \frac{1}{\sqrt{8}} \right)$
  \\
IADCT~\cite{Potluri2013} &
$ \left[\begin{smallmatrix}
1 & 1 & 1 & 1 & 1 & 1 & 1 & 1 \\
0 & 1 & 0 & 0 & 0 & 0 &-1 & 0 \\
1 & 0 & 0 &-1 &-1 & 0 & 0 & 1 \\
1 & 0 & 0 & 0 & 0 & 0 & 0 &-1 \\
1 &-1 &-1 & 1 & 1 &-1 &-1 & 1 \\
0 & 0 & 0 & 1 &-1 & 0 & 0 & 0 \\
0 &-1 & 1 & 0 & 0 & 1 &-1 & 0 \\
0 & 0 & 1 & 0 & 0 &-1 & 0 & 0 \\
\end{smallmatrix}\right]$
 & $\operatorname{diag} \left(\frac{1}{\sqrt{8}}, \frac{1}{\sqrt{2}},\frac{1}{2}, \frac{1}{\sqrt{2}}, \frac{1}{\sqrt{8}}, \frac{1}{\sqrt{2}}, \frac{1}{2}, \frac{1}{\sqrt{2}} \right)$
\\
\bottomrule
\end{tabular}
\end{table*}

\section{Multidimensional Approximate DCT}
\label{sec:3d_dct_approx}

Although
widely examined as a tool for image/video compression~\cite{haweel2001,lengwehasatit2004scalable,cb2011,bc2012,bas2008,bas2009,bas2013, Potluri2013,cintra2011integer,Cintra2014-sigpro},
DCT approximations lack
a formal mathematical definition
for general multidimensional case.
In the present section,
we focus on deriving algebraic expressions based on
the above-mentioned tensor analysis.
As a consequence,
we propose
approximate 3D methods
based on
state-of-the-art 1D DCT approximations.
We also evaluate
the arithmetic complexity of
the general multidimensional DCT approximation
with
emphasis on the 3D case.

\subsection{Mathematical Definition}
\label{sec_3d_dct_approx_math_def}

We aim at approximating
the multidimensional DCT
based on the vector bases
of
1D~DCT
approximations~\cite{haweel2001,lengwehasatit2004scalable, cb2011, bc2012, bas2008, bas2009, bas2013,Potluri2013}.
Such associate basis vectors
often lack closed-form expressions,
being generally derived from
numerical computation and/or brute-force search~\cite{Cintra2014-sigpro}.
Therefore,
in general,
simple analytic expressions similar
to~\eqref{1d_dct_analit},
\eqref{2d_dct_analit},
and~\eqref{3d_dct_analit} are not available.
However,
we can derive
multidimensional
DCT approximations
by means of the high-order tensor formalism.
We first focus on the three dimensional case.
Let $\mathcal{T} \in \mathbb{F}^{N \times N \times N}$
be a third-order tensor representing a given input discrete signal.
The associate transform-domain output
third-order tensor
$\mathcal{Y} \in \mathbb{F}^{N \times N \times N}$
is given by:
\begin{align}
\begin{split}
\mathcal{Y} &
\triangleq
\mathcal{T} \times_1
\hat{\mathbf{C}}_N  \times_2
\hat{\mathbf{C}}_N  \times_3
\hat{\mathbf{C}}_N
\\
&=
\mathcal{T} \times_1
(\mathbf{S}_N \cdot \mathbf{T}_N )
\times_2
(\mathbf{S}_N \cdot \mathbf{T}_N )
\times_3
(\mathbf{S}_N \cdot \mathbf{T}_N )
.
\end{split}
\label{3d_dct_approx1}
\end{align}
From the properties expressed in~\eqref{porp_comut} and~\eqref{porp_dist},
we can recast~\eqref{3d_dct_approx1} according to:
\begin{align}
\mathcal{Y} =
\mathcal{T}
\times_1
\mathbf{T}_N
\times_2
\mathbf{T}_N
\times_3
\mathbf{T}_N
\times_1
\mathbf{S}_N
\times_2
\mathbf{S}_N
\times_3
\mathbf{S}_N
.
\label{3d_dct_approx2}
\end{align}
Therefore,
the 3D approximate DCT can be computed by first computing
the $i$-mode products
by the low complexity matrices $\mathbf{T}_N$.
The operations involving
the diagonal matrix~$\mathbf{S}_N$
can be
efficiently combined and computed separately.

The inverse transformation is related to
the inverse approximate DCT matrix
given in~\eqref{inverse_approx_matrix}.
Considering expressions~\eqref{porp_comut},~\eqref{porp_dist},~\eqref{porp_ident}, and~\eqref{3d_dct_approx1},
the inverse 3D~DCT approximation
is given by:
\begin{align}
\mathcal{T} &=
\mathcal{Y} \times_1
\hat{\mathbf{C}}_N^{-1}  \times_2
\hat{\mathbf{C}}_N^{-1}  \times_3
\hat{\mathbf{C}}_N^{-1}.
\label{3d_idct_approx_1}
\end{align}

We can extend
the above expressions
to
the multidimensional case
to
derive
$R$-dimensional DCT approximations
of size
$N_1 \times N_2 \times \cdots \times N_R$.
Let
$\left\{ \hat{\mathbf{C}}_{N_i} \right\}_{i=1}^{R}$
be a collection of
$N_i$-point DCT approximations matrices,
where
$\hat{\mathbf{C}}_{N_i} = \mathbf{S}_{N_i} \cdot \mathbf{T}_{N_i}$
for $i=1,2,\ldots,R$,
as described in Section~\ref{sec_sub:DCT_approximations}.
Let
$\mathcal{T} \in \mathbb{F}^{N_1 \times N_2 \times \cdots \times N_R}$
be an $R$th-order tensor representing an input data array.
The approximate transform-domain output data is defined by:
\begin{align}
\begin{split}
\mathcal{Y}
\triangleq
&
\mathcal{T} \times_1
\hat{\mathbf{C}}_{N_1}
\times_2
\hat{\mathbf{C}}_{N_2}
\times_3
\cdots
\times_R
\hat{\mathbf{C}}_{N_R}
\\
= &
\mathcal{T}
\times_1
(\mathbf{S}_{N_1} \cdot \mathbf{T}_{N_1} )
\times_2
(\mathbf{S}_{N_2} \cdot \mathbf{T}_{N_2} )
\times_3
\cdots
\\ &
\times_R
(\mathbf{S}_{N_R}
\cdot
\mathbf{T}_{N_R} )
\\
= &
\mathcal{T}
\times_1
\mathbf{T}_{N_1}
\times_2
\mathbf{T}_{N_2}
\times_3
\cdots
\times_R
\mathbf{T}_{N_R}
\\ &
\times_1
\mathbf{S}_{N_1}
\times_2
\mathbf{S}_{N_2}
\times_3
\cdots
\times_R
\mathbf{S}_{N_R}
.
\end{split}
\label{multd_dct_approx}
\end{align}
Notice that there can be different 1D DCT approximations
for a fixed blocklength.
Thus,
if $N_i = N_j$, for $i\neq j$,
the approximations
for the $i$th and $j$th dimension may not necessarily
be the same---although their blocklengths are the same.
However,
selecting identical approximations
for the identical blocklength
seems to be a natural choice.
The inverse multidimensional transformation is computed
according to the following expression:
\begin{align}
\mathcal{Y} =
\mathcal{T}
\times_1
\hat{\mathbf{C}}_{N_1}^{-1}
\times_2
\hat{\mathbf{C}}_{N_2}^{-1}
\times_3
\cdots
\times_R
\hat{\mathbf{C}}_{N_R}^{-1},
\label{multd_idct_approx_1}
\end{align}
where
$\hat{\mathbf{C}}_{N_i}^{-1}$
is derived from~\eqref{inverse_approx_matrix}.

\subsection{Complexity Assessment}
\label{sec_complexity_assessment}

\begin{table*}[t]
\centering
\caption{Coding efficiency and computational complexity assessment}
\label{table_complexity}
\begin{tabular}{l cc | ccc | ccc}
\toprule
 &
\multicolumn{2}{c}{Efficiency} &
\multicolumn{3}{c}{1D Complexity} &
\multicolumn{3}{c}{3D Complexity}
\\
\cmidrule {2-9}
Method &
$C_g$ (dB) & $\eta~(\%)$  &
Mult. & Add. & Shift & Mult. & Add. & Shift
 \\
\midrule
DCT (by definition)~\cite{Ahmed1974}  &
8.83 & 93.99 &
64 & 56 & 0 & 12288 & 10752 & 0
\\
Loeffler DCT algorithm~\cite{Chen1977} &
8.83 & 93.99 &
11 & 29 & 0 & 2112 & 5568 & 0 \\
Chen DCT algorithm~\cite{Chen1977} &
8.83 & 93.99 &
16 & 26 & 0 & 3072 & 4992 & 0 \\
3D~VR~DIF algorithm~\cite{Boussakta2004} &
8.83 & 93.99 &
-- & -- & -- & 1344 & 5568 & 0 \\
\color{black}
SDCT~\cite{haweel2001} &
6.03 & 82.62 &
0 & 24 & 0 & 0 & 4608 & 0 \\
LODCT~\cite{lengwehasatit2004scalable} &
8.39 & 88.70 &
0 & 24 & 2 & 0 & 4608 & 384 \\
RDCT~\cite{cb2011} &
8.18 & 87.43 &
0 & 22 & 0 & 0 & 4224 & 0 \\
MRDCT~\cite{bc2012} &
7.33 & 80.90 &
0 & 14 & 0 & 0 & 2688 & 0 \\
BAS-2008~\cite{bas2008} &
8.12 & 86.86 &
0 & 18 & 2 & 0 & 3456 & 384 \\
BAS-2009~\cite{bas2009} &
7.91 & 85.38 &
0 & 18 & 0 & 0 & 3456 & 0 \\
BAS-2013~\cite{bas2013,kouadria2013low} &
7.95 & 85.31 &
0 & 24 & 0 & 0 & 4608 & 0\\
IADCT~\cite{Potluri2013} &
7.33 & 80.90 &
0 & 14 & 0 & 0 & 2688 & 0 \\
\bottomrule
\end{tabular}
\end{table*}

Due to the kernel separability property~\cite{Kok1997},
the exact and approximate multidimensional DCT
can be computed by successive instantiations of the 1D~DCT~\cite{Boussakta2004,Servais1997,lengwehasatit2004scalable}.
Consequently,
a fast algorithm for the approximate 1D~DCT can be applied for higher dimensions.
In~\eqref{multd_dct_approx},
$R$~different $i$-mode products are employed.
Thus,
applying~\eqref{general_tensor_prod}
to~\eqref{multd_dct_approx}
for a specific
dimension~$i$,
we obtain an expansion
with
$N_1\cdot N_2 \cdots N_{i-1} \cdot N_{i+1} \cdots N_R$
free indices
and
$i$-mode products.
The same reasoning
can be applied to
the remaining dimensions.

Let
$\mathcal{A}_{\text{1D}}\left(\hat{\mathbf{C}}_{N_i} \right)$
be the
arithmetic complexity of
$\hat{\mathbf{C}}_{N_i}$.
Such complexity
encompasses
multiplicative,
additive,
and
bit-shifting
costs,
which
depends on the
considered
fast algorithm.
Then,
the arithmetic complexity
for the $R$-dimensional case is generally given by:
\begin{align}
\label{multd_arithmetic_complexity}
\mathcal{A}_{R\text{D}}
\left(
\hat{\mathbf{C}}_{N_1},
\hat{\mathbf{C}}_{N_2},
\ldots,
\hat{\mathbf{C}}_{N_R}
\right)
=
 \sum_{i=1}^{R}
\Omega^{(i)}_R
\cdot
\mathcal{A}_{\text{1D}}\left(\hat{\mathbf{C}}_{N_i} \right)
,
\end{align}
where
\begin{align}
\Omega^{(i)}_R
\triangleq
\begin{cases}
1, & \text{if } R=1, \\
\prod\limits_{\substack{j=1 \\ j\neq i}}^R N_j, & \text{otherwise.}
\end{cases}
\end{align}

For the 1D case,
the right-hand side of
\eqref{multd_arithmetic_complexity} becomes
$\mathcal{A}_{\text{1D}}\left(\hat{\mathbf{C}}_{N_1} \right)$.
If
(i)~$N_1 = N_2 = \cdots = N_R \triangleq N$,
$R \geq 2$,
and
(ii)~the same approximate matrix~$\hat{\mathbf{C}}_{N}$
is considered in all dimensions,
then
the arithmetic complexity
is given by:
\begin{align}
\label{multd_arithmetic_complexity_part_case}
\mathcal{A}_{R\text{D}}
\left(
\hat{\mathbf{C}}_{N}
\right)
&=
R\cdot N^{R-1} \cdot
\mathcal{A}_{\text{1D}}\left(\hat{\mathbf{C}}_{N} \right).
\end{align}
Although our approach is general
and suitable for any blocklength,
we focus our attention
in $N=8$.
Indeed,
this particular blocklength
is relevant in a significant number
of practical contexts~\cite{hevc,Lecuire2012,mpeg2,rao1996techniques,Wallace1992}.
Thus we can benefit of state-of-the-art methods.
\color{black}

Table~\ref{table_complexity}
shows the computational costs
for $R=3$
for several proposed 3D~DCT approximations.
Transformations were based on the
discussed 1D~DCT approximations
in~Table~\ref{tab_dct_approx}
and computed according to~\eqref{3d_dct_approx2}.
All approximate methods present null multiplicative complexity.
For reference,
we also
included
the computational cost
of the exact DCT
evaluated
according to
(i)~its definition,
(ii)~to
the Loeffler DCT algorithm~\cite{Loeffler1989},
and
(iii)~to the Chen DCT algorithm~\cite{Chen1977}.
The Loeffler DCT algorithm
achieves the theoretical minimum
multiplicative complexity for 1D case~\cite{Heideman1988}
and
the Chen DCT algorithm
is employed in the HEVC standard~\cite{hevc}.
We also included the costs
of
the 3D~VR~DIF algorithm~\cite{Boussakta2004},
which computes the exact 3D~DCT directly
without the row-column approach and requires less
multiplications operations than the Loeffler DCT algorithm for 3D case.
Table~\ref{table_complexity}
also
displays
popular
coding performance measures~\cite[p.~163]{britanak2007discrete}:
(i)~coding gain~$C_g$; and
(ii)~transform efficiency~$\eta$.
\color{black}
Table~\ref{table_complexity_reduction}
shows
the percent
reductions
in the
coding performance measurements and
\color{black}
computational complexity
for each 3D~method
when compared with the
exact 3D~DCT,
considering the 3D~VR~DIF direct algorithm.
The total complexity reduction
was obtained considering
the sum of all arithmetic operations in Table~\ref{table_complexity}.
The percent reductions
in complexity
offered
by the proposed approximations
exceeds
the percent reductions
in performances.
This fact suggests a favorable trade-off.
\color{black}
The 3D~MRDCT and 3D~IADCT methods
equally
present
61.1\% and 51.7\%
reductions
in total arithmetic cost
and
number of additions,
respectively,
when compared with
the 3D~VR~DIF algorithm.
The 3D~LODCT shows
the smallest performance degradation:
4.9\% and 5.6\%
reductions
in
coding gain and transform efficiency reduction,
respectively,
at a total computational saving of 27.8\%.
\color{black}
\subsection{Trade-off Analysis}
\label{sec_trade_off}

The overall performance
of a particular
approximate 3D~DCT
in a specific contexts
depends on
a large number of factors~\cite{briggs1995dft}.
A variety
of trade-off effects
are present,
being a very hard task to precisely quantify
them~\cite{Lecuire2012,cintra2015energy}.
However,
an initial trade-off
analysis
can be obtained by means
of a combined figure of merit~$f$
that
takes into
consideration
computational complexity
and
coding performance,
which
are two major metrics in the field~\cite{Blahut2010,britanak2007discrete}.
We propose the following convex combination
as the figure of merit
for assessing the discussed
3D~DCT approximations:
\begin{equation}
\label{cost_function}
\begin{split}
f
\triangleq
&
\gamma
\cdot
(\text{normalized computational complexity})
\\
&
+
(1-\gamma)
\cdot
(\text{normalized performance}),
\end{split}
\end{equation}
where
$\gamma \in [0,1]$ is a weighting factor
and
the normalization is taken
to ensure the variables are comparable.
Quantity~$\gamma$~adjusts the importance of each
component---computational complexity
and performance---according to
the given context.
Such type of figure of merit
is often found in optimization literature
referred to as `cost function'~\cite{ehrgott2005multicriteria}.

For the computational complexity metric in~\eqref{cost_function},
we adopt the arithmetic cost
given by
the
weighted sum of the multiplicative, additive, and bit-shift complexities.
Let
$M$, $A$, and $S$
be the number of
multiplications,
additions, and
bit-shifting operations,
respectively;
and
$w_m$,
$w_a$, and
$w_s$ their respective
weighting costs.
Then,
we define:
\begin{align}
\begin{split}
(\text{arithmetic cost})
& \triangleq
w_m \cdot
M
+
w_a \cdot
A
+
w_s \cdot
S \\
&= w_m
\cdot \left(
M
+
\frac{w_a}{w_m} \cdot
A
+
\frac{w_s}{w_m} \cdot
S
\right)
\\
&= w_m \cdot
\left(
M
+
\beta \cdot
A
+
\beta' \cdot
S
\right),
\end{split}
\label{arith_cost}
\end{align}
where
$\beta = w_a/w_m$ and $\beta'=w_s/w_m$.
However,
because bit-shifting operations
require considerably lower energy consumption
and hardware resources
when compared with
additions or multiplications~\cite{Blahut2010,malvar2003low,Potluri2013},
we have that
$w_m \gg w_s$;
therefore
we obtain $\beta'\approx 0$.
Moreover,
because~\eqref{cost_function}
takes
the normalized complexity
(relative to the largest measured value
among all methods),
the term $w_m$
has no practical effect.
Hereafter,
for simplicity,
we can consider $w_m = 1$.
Consequently,
we obtain:
\begin{align}
(\text{arithmetic cost})
&
\approx
M
+
\beta
A
.
\label{arith_cost_simp}
\end{align}
Estimating the actual value for
$\beta$
is not an easy task,
as discussed in~\cite[p.~395]{briggs1995dft}.
However,
it is a well-known fact that
multiplying
is a more complex operation
than
adding
($w_m > w_a$)
both in
hardware and software implementations~\cite{Blahut2010}.
Thus,
we have that
$\beta < 1$.
Moreover,
computational systems
tend to
satisfy
$w_a \ll w_m$;
thus
low values of $\beta$
are expected in
practice.
In view of the above,
we adopted $\beta \in [0,1]$
in our analysis.
In its turn,
for the performance metric in~\eqref{cost_function},
we adopted
the negative of the coding gain,
so the figure of merit
decreases in value
as the coding performance
improves.
Transform efficiency
was not included
because
it is correlated to coding gain.

\begin{figure}[t]
 \centering
        \includegraphics[scale=1]{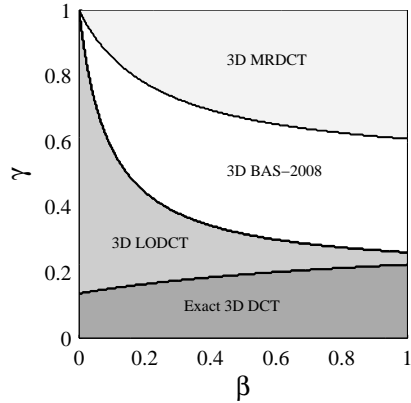}
\caption{Optimum 3D transformations for different cost function parameters.}
\label{fig_tradeoff}
\end{figure}

We submitted
all discussed 3D~DCT approximations
and the exact 3D~DCT
considering the VR~DIF algorithm
to the proposed figure of merit
for all values of
$\gamma$
and
$\beta$.
In~\figurename~\ref{fig_tradeoff},
we
labeled regions
according
to the particular 3D transformation
that
excels
in that particular plot area
(combinations of $\gamma$ and $\beta$).
Low values of~$\gamma$
emphasize coding performance;
thus---as expected---the exact 3D~DCT
tends to be the optimum choice.
From $\gamma \approx 0.2$,
approximations surpass the exact computation.
For
middle-low,
middle-high, and
high
$\gamma$ values,
the 3D~LODCT,
the 3D~BAS-2008, and
the 3D~MRDCT
are the optimized methods,
respectively.
Analyzing the $\beta$-axis,
for low $\beta$ values,
which
emphasizes the
multiplicative over arithmetic complexity,
the 3D~LODCT tend to outperform
the competing approximations.
By incrementing $\beta$,
which amplifies the importance
of the additive complexity for the overall computational cost,
the 3D~BAS-2008 and the 3D~MRDCT
become more relevant;
occupying a larger area in the plot.
Also,
the area where
the exact 3D~DCT is the best method slightly grows
as $\beta$ increases.
The proposed approximate 3D methods
occupy
a vastly larger area of the plot;
outperforming
the exact 3D~DCT method
in realistic scenarios.

\color{black}

\begin{table}[t]
\centering
\caption{Coding efficiency and computational complexity percent reduction relative to the exact 3D~DCT (considering 3D~VR~DIF direct algorithm)}
\label{table_complexity_reduction}
\begin{tabular}{l cc | ccc}
\toprule
 &
\multicolumn{2}{c}{Efficiency} &
\multicolumn{3}{c}{3D Complexity}
\\
\cmidrule {2-6}
Method & $C_g$ & $\eta$ & Mult. & Add. & Total
 \\
\midrule
3D~SDCT &
31.7 \% & 12.1 \% &
100 \% & 17.2 \% & 33.3 \%  \\
3D~LODCT &
\textbf{4.9} \% & \textbf{5.6} \% &
100 \%& 17.2 \% & 27.8 \%  \\
3D~RDCT &
7.3 \% & 7.0 \% &
100 \%& 24.1 \% & 38.9 \%  \\
3D~MRDCT &
16.9 \% & 13.9 \% &
100 \%& \textbf{51.7 \%} & \textbf{61.1 \%}  \\
3D~BAS-2008 &
8.0 \% & 7.6 \% &
100 \%& 37.9 \% & 44.4 \%  \\
3D~BAS-2009 &
10.4 \% & 9.2 \% &
100 \%& 37.9 \% & 50.0 \%  \\
3D~BAS-2013 &
 10.0 \%& 9.2 \% &
100 \%& 17.2 \% & 33.3 \% \\
3D~IADCT &
16.9 \% &  13.9 \% &
100 \%& \textbf{51.7 \%} & \textbf{61.1 \% } \\
\bottomrule
\end{tabular}
\end{table}
\color{black}

\section{Interframe Video Coding Based on 3D~DCT Approximations}
\label{sec:interframe_video_coding}

\begin{figure*}[t]
 \begin{subfigure}[b]{1\linewidth}
 \centering
        \includegraphics[scale=1]{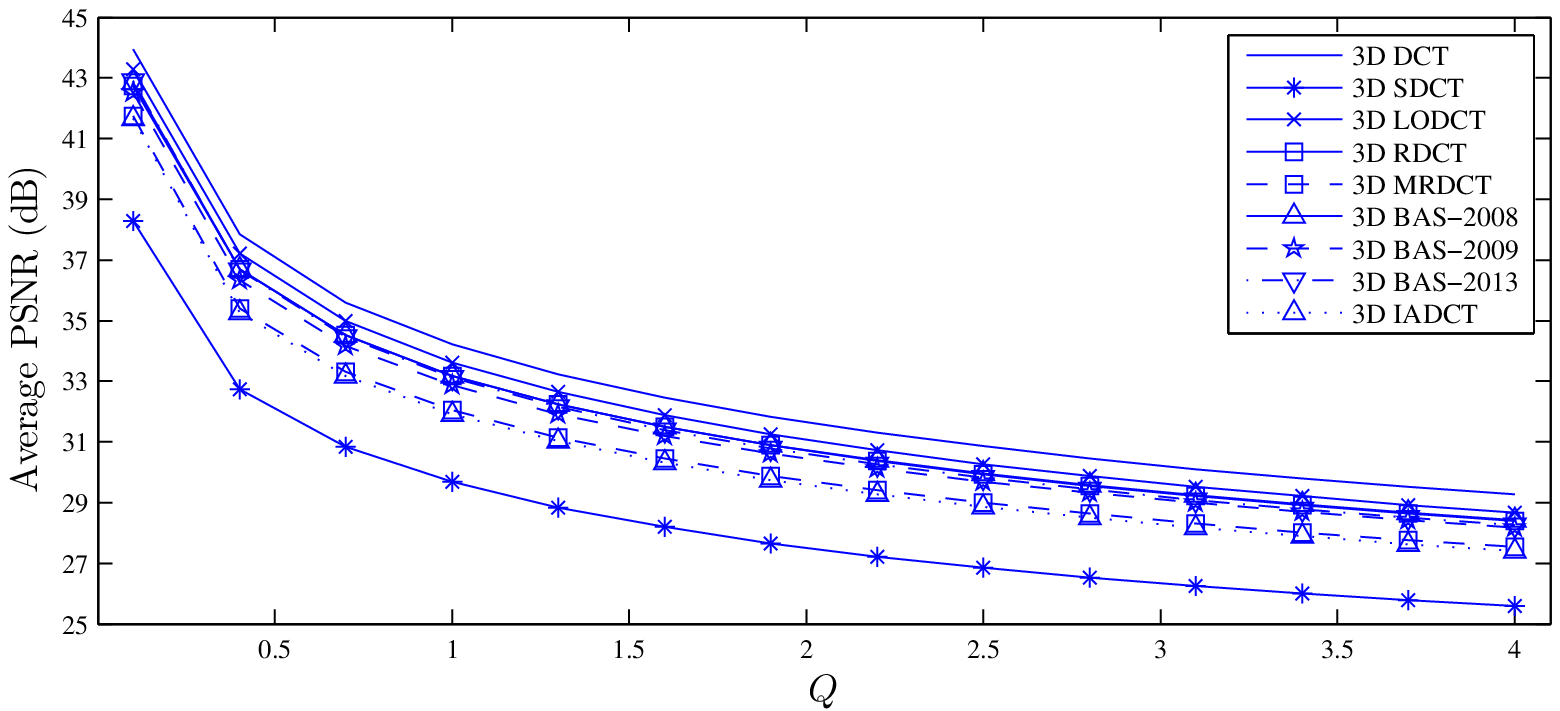}
        \caption{PSNR}
        \label{fig_psnr}
 \end{subfigure} \\
 \begin{subfigure}[b]{1\linewidth}
 \centering
        \includegraphics[scale=1]{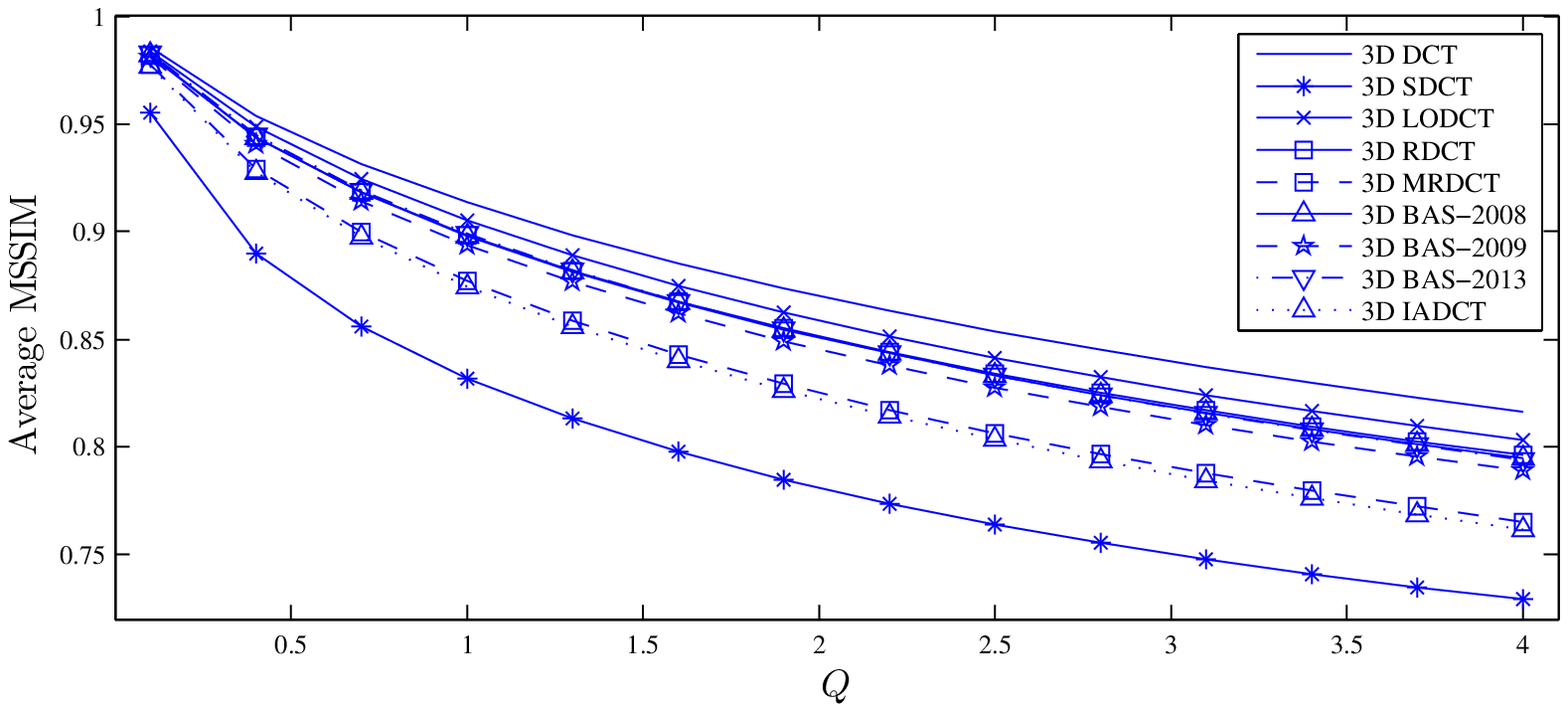}
        \caption{MSSIM}
      \label{fig_mse}
 \end{subfigure}
\caption{Video compression performance measures.}
\label{fig_perfomance}
\end{figure*}

This section
introduces
low-complexity interframe
video coding
schemes~\cite{natarajan1977interframe,Servais1997,abousleman1995compression,lai2002video,li2007multiview}
equipped
with the discussed 3D~DCT approximations
(cf.~Table~\ref{tab_dct_approx}
and
\eqref{3d_dct_approx2}).
We embed the 3D~DCT approximations
into
a three dimensional block transform coding system based on 3D~DCT~\cite{natarajan1977interframe,Servais1997,abousleman1995compression,lai2002video,li2007multiview}.
We also propose a method to modify
a given 3D~quantization volume in order
to avoid extra computation
of such 3D approximate transforms.
Then
we submit
a set of widely employed video sequences
to
the
modified video coding system
to evaluate output video quality
relative to the original coding scheme.

\subsection{Modified Quantization Procedure for Video Compression}
\label{sec:quantization}

In image compression schemes,
the quantization step
plays a fundamental role
since
it
non-linearly re-scales or discards
transform-domain components according
to
(i)~their respective importance
to perceived visual quality~\cite[p.~155]{bhaskaran1997}
or
(ii) data correlational properties~\cite[p.~156]{bhaskaran1997}.
In
2D~DCT coding,
the quantization procedure depends on quantization
tables (or matrices)~\cite{Rao2001,bhaskaran1997}
prescribed by adopted standards,
such as
MPEG~\cite{Wallace1992,mpeg2},
H.261~\cite{h261}
H.263~\cite{h263},
H.264~\cite{h264},
and
HEVC~\cite{hevc}.

As detailed in~\cite{britanak2007discrete,tran2000bindct,Liang2001,lengwehasatit2004scalable},
the diagonal matrix of the approximate DCT
(cf.~\eqref{equation-S-T})
can be merged into the quantization tables,
eliminating the need for
additional computation implied by the diagonal elements.
However,
for the 3D~DCT based coding,
such tables are not adequate
because
they are 2D arrays in nature~\cite{lee1997quantization,Chan1997,bozinovic2003scan,bozinovic2005}.
Instead of a quantization table,
a quantization volume is required.
In~\cite{lee1997quantization,Chan1997,bozinovic2003scan},
methods to generate 3D quantization volumes are proposed.
In this section,
we aim at proposing
a method for
embedding
the diagonal matrices
described in~\eqref{3d_dct_approx2}
into a given quantization volume.

Let $\mathcal{Q} \in \mathbb{F}^{N \times N \times N}$
be a previously designed 3D quantization volume,
whose entries are given by
$q[k_1,k_2,k_3]$, for $k_1,k_2,k_3 = 0,1,\ldots,N-1$.
The quantization step
performs
the following computation~\cite{lee1997quantization,Chan1997}:
\begin{align}
\begin{split}
\tilde{y}[k_1,k_2,k_3] = \operatorname{round} \left(\frac{y[k_1,k_2,k_3]}{q[k_1,k_2,k_3]}
\right),
\\
k_1,k_2,k_3 = 0,1,\ldots, N-1,
\end{split}
\label{quantiz_usual}
\end{align}
where
$y[k_1,k_2,k_3]$
are transform-domain coefficients
according to~\eqref{3d_dct_approx2}
and
$\tilde{y}[k_1,k_2,k_3]$
are the transform-domain quantized coefficients.
The dequantization process is defined by
$\hat{y}[k_1,k_2,k_3] = \tilde{y}[k_1,k_2,k_3] \cdot q[k_1,k_2,k_3]$,
where
$\hat{y}[k_1,k_2,k_3]$
is an estimative of
${y}[k_1,k_2,k_3]$~\cite{bhaskaran1997}.

Now,
we split the computation of~\eqref{3d_dct_approx2}
into two steps:
(i)~the $i$-mode products involving the low-complexity matrix
$\mathbf{T}$,
given by:
\begin{align}
\mathcal{A} =
\mathcal{T}
\times_1
\mathbf{T}_N
\times_2
\mathbf{T}_N
\times_3
\mathbf{T}_N,
\label{tensor_A}
\end{align}
and
(ii)~the $i$-mode products requiring the diagonal matrix~$\mathbf{S}_N$.
In the Appendix,
we demonstrate the following expression:
\begin{align}
\begin{split}
y[k_1, k_2, k_3] &=
a[k_1, k_2, k_3]
\cdot
d_{k_1}
\cdot
d_{k_2}
\cdot
d_{k_3}
,
\end{split}
\label{tensor_y}
\end{align}
where
$a[k_1, k_2, k_3]$
are tensor $\mathcal{A}$ entries
given in~\eqref{tensor_A}
and
$d_k$ is the $k$th diagonal element of~$\mathbf{S}_N$.
Applying~\eqref{tensor_y} into~\eqref{quantiz_usual},
we obtain:
\begin{align}
\label{quantiz_usual_2}
\begin{split}
\tilde{y}[k_1,k_2,k_3]
=
\operatorname{round}
\left(
\frac{a[k_1, k_2, k_3]
\cdot
d_{k_1}
\cdot
d_{k_2}
\cdot
d_{k_3}}
{q[k_1,k_2,k_3]}
\right)
,
\\
k_1,k_2,k_3 = 0,1,\ldots, N-1.
\end{split}
\end{align}

We propose
a new quantization volume
$\mathcal{Q}^* \in \mathbb{F}^{N\times N \times N}$,
whose entries are computed by:
\begin{align}
q^*[k_1,k_2,k_3]
\triangleq
\frac{q[k_1,k_2,k_3]}{
d_{k_1}
\cdot
d_{k_2}
\cdot
d_{k_3}
}
.
\label{q_modified}
\end{align}
Replacing~\eqref{q_modified}
in~\eqref{quantiz_usual_2},
we obtain
the modified quantization step
as follows:
\begin{align}
\begin{split}
\tilde{y}[k_1,k_2,k_3] =
\operatorname{round}
\left(\frac{a[k_1,k_2,k_3]}{q^*[k_1,k_2,k_3]} \right),
\\
k_1,k_2,k_3 = 0,1,\ldots, N-1.
\end{split}
\label{quantiz_modified}
\end{align}
Notice that only the tensor $\mathcal{A}$
is required to be computed.
It is employed as
the input data
to
the
modified quantization step~\eqref{quantiz_modified}.
Because the
computation of~$\mathcal{A}$
demands
only
the low-complexity transformation~$\mathbf{T}_N$
(see~\eqref{tensor_A}),
the
computational overhead
imposed by diagonal matrix
is eliminated.
The modified dequantization process
can be obtained by a similar procedure
and
it is furnished by:
$\hat{y}[k_1,k_2,k_3] = \tilde{y}[k_1,k_2,k_3] \cdot q^\star[k_1,k_2,k_3]$,
where
\begin{align}
q^\star[k_1,k_2,k_3]
\triangleq
{q[k_1,k_2,k_3]}
\cdot
{
d_{k_1}
\cdot
d_{k_2}
\cdot
d_{k_3}
}
.
\label{qinv_modified}
\end{align}

\subsection{Video Compression Simulation}
\label{sec:video_compression}

A
video compression scheme
based on~\cite{natarajan1977interframe,Servais1997,abousleman1995compression,lai2002video,li2007multiview}
was considered.
Nine standard CIF~\cite{fitzek2001mpeg}
YUV video sequences available
at~\cite{xiph_database}
were chosen
as third-order tensors of size $N_1 \times N_2 \times N_3$.
The CIF video sequences
present
$N_1 = 352$ and $N_2=288$;
and we selected 296~consecutive frames ($N_3=296$).
Each video tensor was divided
into $8 \times 8 \times 8$ smaller tensors,
which were applied to
the 3D~transformation defined in~\eqref{3d_dct_approx2},
and then submitted to the quantization stage.
The
1D~DCT
approximations shown in
Table~\ref{tab_dct_approx}
were considered
to derive the 3D approximate methods.
In addition,
the exact 3D~DCT was also included in the experiments
for comparison.
The quantization volume suggested in~\cite{lee1997quantization}
was applied to the modified quantization procedure,
as described in Section~\ref{sec:quantization}.
The procedure was simulated
at quality factors $Q$
ranging in
$Q \in [0.1, 4.0]$~\cite{bhaskaran1997}.
The inverse procedure was applied
to reconstruct the video sequence.
The decoding process includes
a modified dequantization and calls to
the inverse approximate 3D~transformation.

To evaluate the data compression performance,
we
employed
the
peak signal to noise ratio~(PSNR)~\cite{bas2008} and
the
structural similarity index~(SSIM)~\cite{Wang2004}
as figures of merit.
The PSNR measure
is related to
the
mean squared error~(MSE)
according
to:
\begin{align}
\operatorname{PSNR}
\triangleq
10 \cdot \log_{10} \left( \frac{255^2}
{\operatorname{MSE}} \right).
\end{align}
Let
$t[n_1,n_2,n_3]$
be the entries of original video tensor
$\mathcal{T}$ and
$\hat{t}[n_1,n_2,n_3]$
be the entries of the recovered video tensor
$\hat{\mathcal{T}}$.
The MSE
of the 3D data
is computed by~\cite{Chan1997}:
\begin{align}
\begin{split}
\operatorname{MSE}
\triangleq
\frac{1}{N_1N_2N_3}
\cdot
\sum_{n_1=0}^{N_1-1}
\sum_{n_2=0}^{N_2-1}
\sum_{n_3=0}^{N_3-1}
e^2[n_1,n_2,n_3]
,
\end{split}
\end{align}
where $e[n_1,n_2,n_3] = t[n_1,n_2,n_3]  - \hat{t}[n_1,n_2,n_3]$.
Since SSIM is defined for still images~\cite{Wang2004},
we computed the SSIM between each original frame
and
its associate compressed frame.
Then,
the mean SSIM~(MSSIM)
taken across frames
was evaluated,
resulting in a performance measure suitable for video sequences.

\begin{figure*}%
\centering
 \begin{subfigure}[t]{0.31\linewidth}
        \includegraphics[scale=0.35]{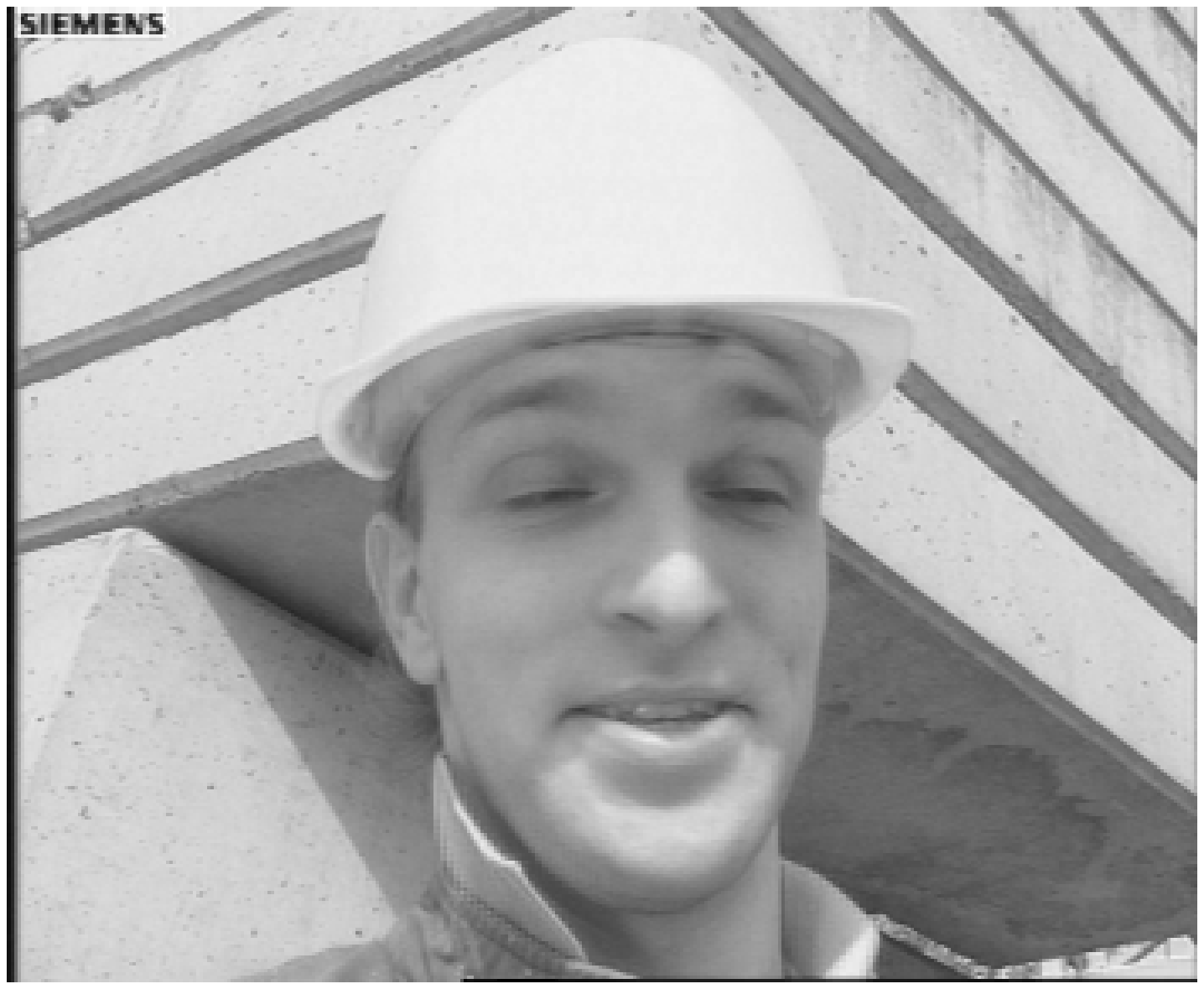}
        \caption{Uncompressed}
 \end{subfigure}
\\
 \begin{subfigure}[t]{0.31\linewidth}
        \includegraphics[scale=0.35]{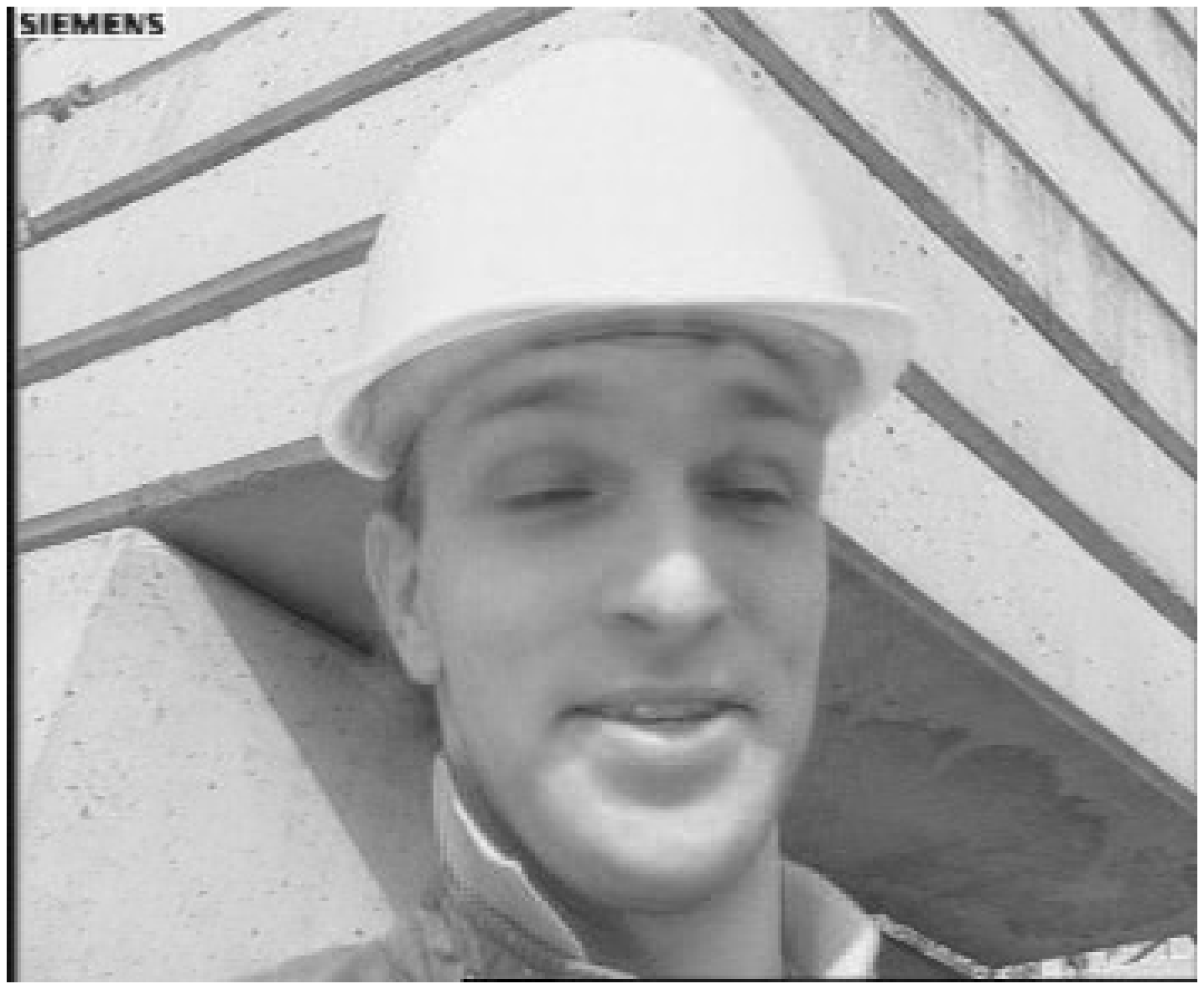}
        \caption{3D~DCT (PSNR=37.8988, MSSIM = 0.9456)}
 \end{subfigure}
 \begin{subfigure}[t]{0.31\linewidth}
        \includegraphics[scale=0.35]{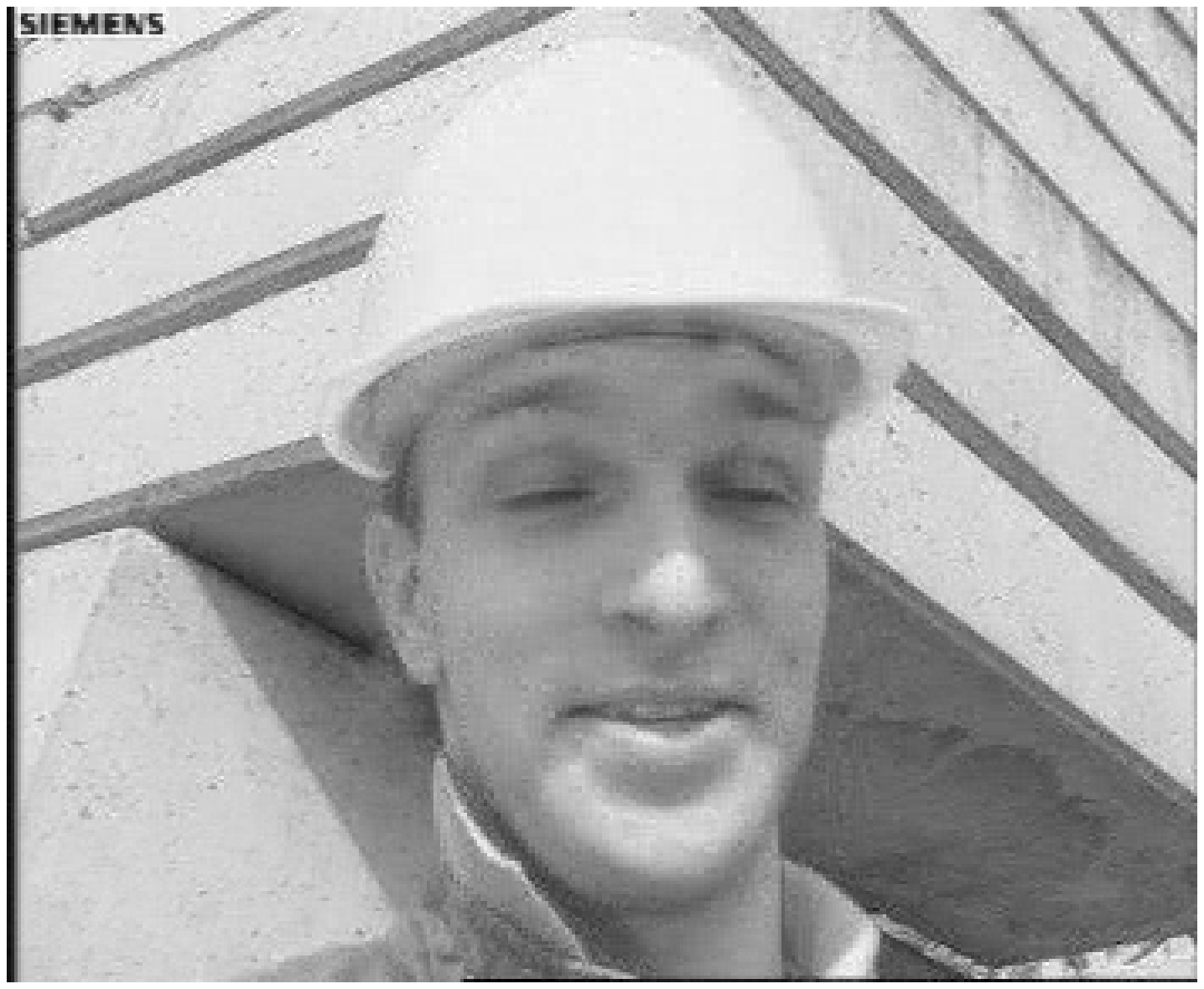}
        \caption{3D~SDCT (PSNR=32.3627, MSSIM = 0.8519)}
 \end{subfigure}
  \begin{subfigure}[t]{0.31\linewidth}
        \includegraphics[scale=0.35]{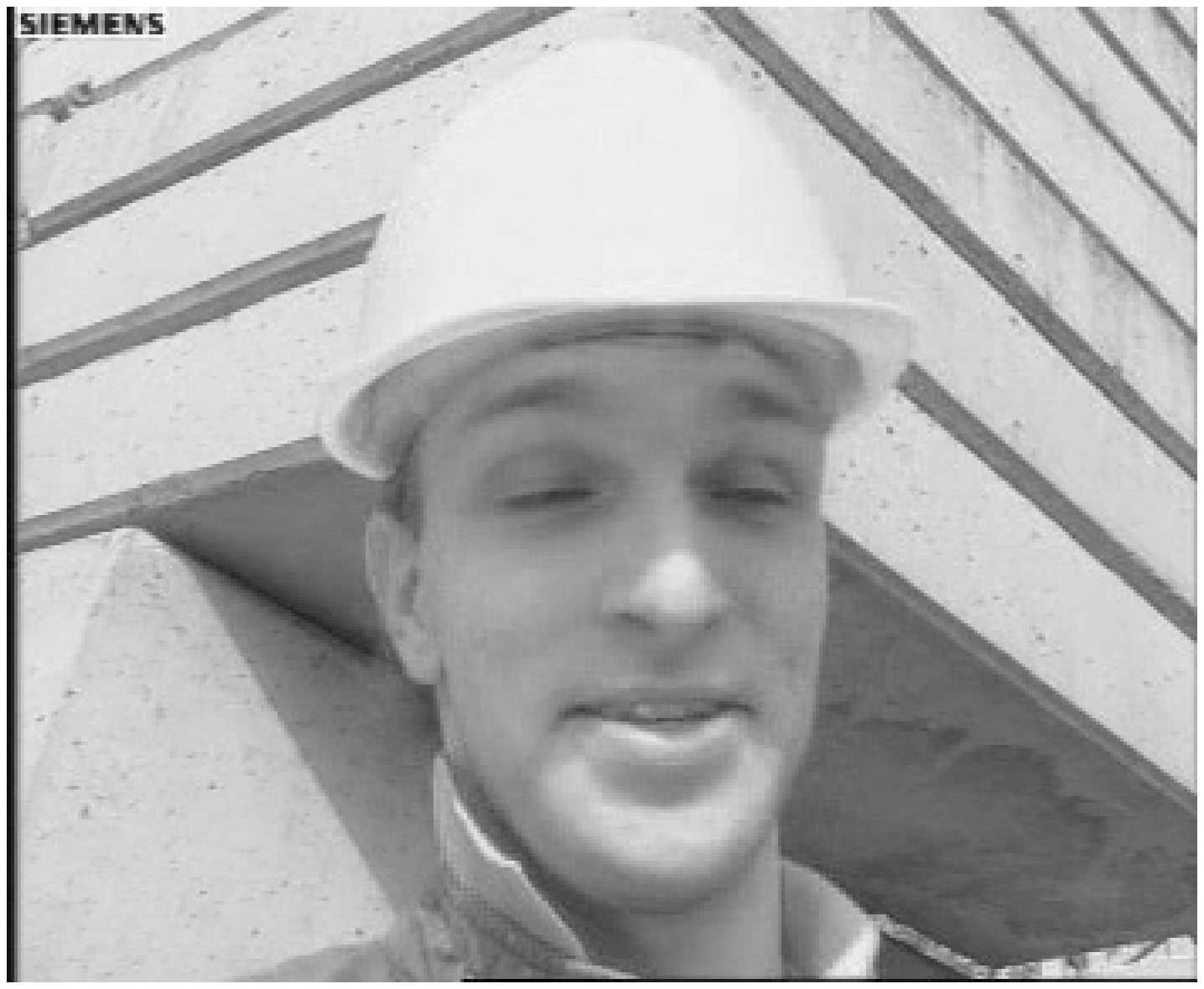}
        \caption{3D~LODCT (PSNR=37.2166, MSSIM = 0.9393)}
 \end{subfigure}
  \begin{subfigure}[t]{0.31\linewidth}
        \includegraphics[scale=0.35]{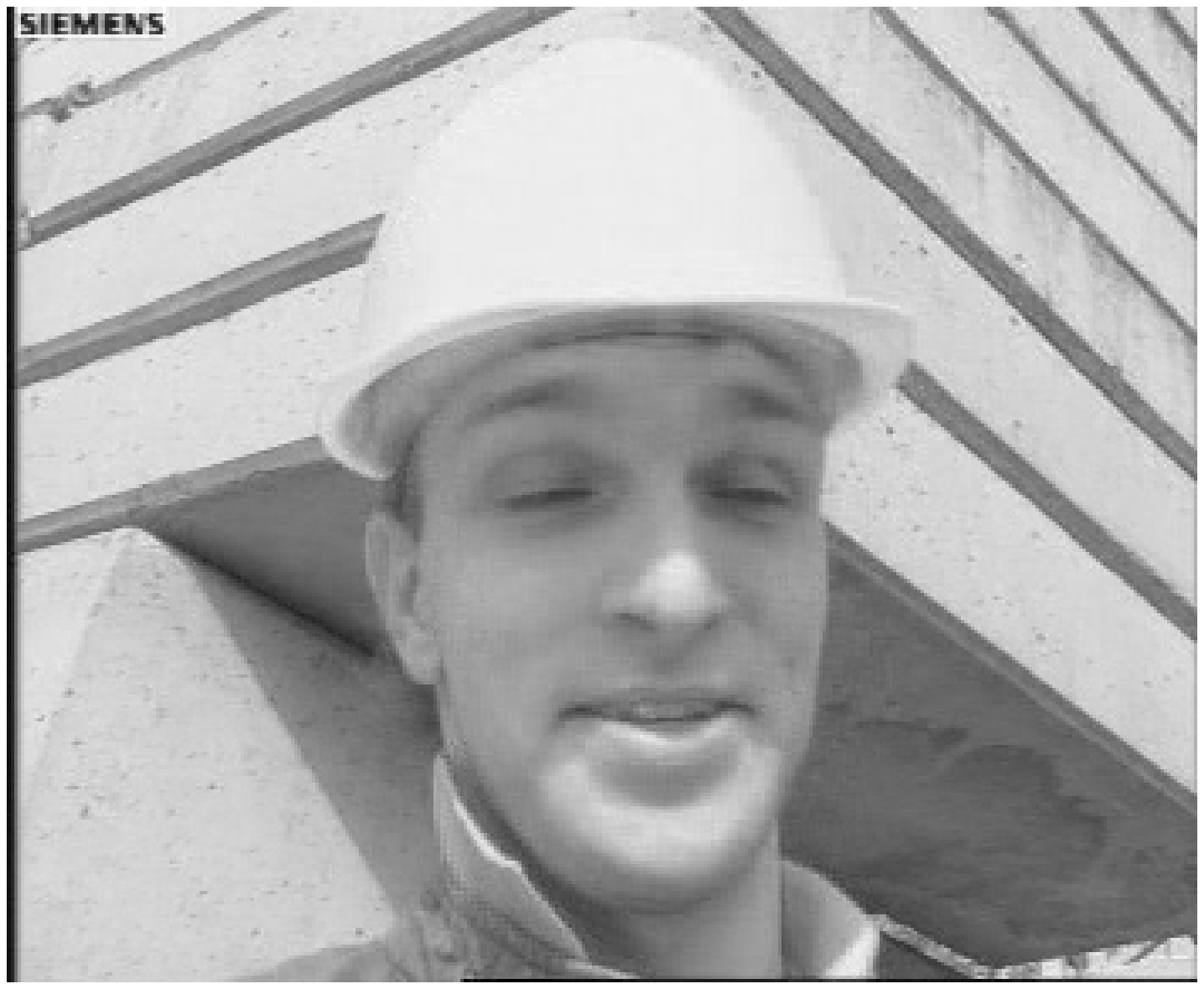}
        \caption{3D~RDCT (PSNR=36.7865, MSSIM = 0.9352)}
 \end{subfigure}
  \begin{subfigure}[t]{0.31\linewidth}
        \includegraphics[scale=0.35]{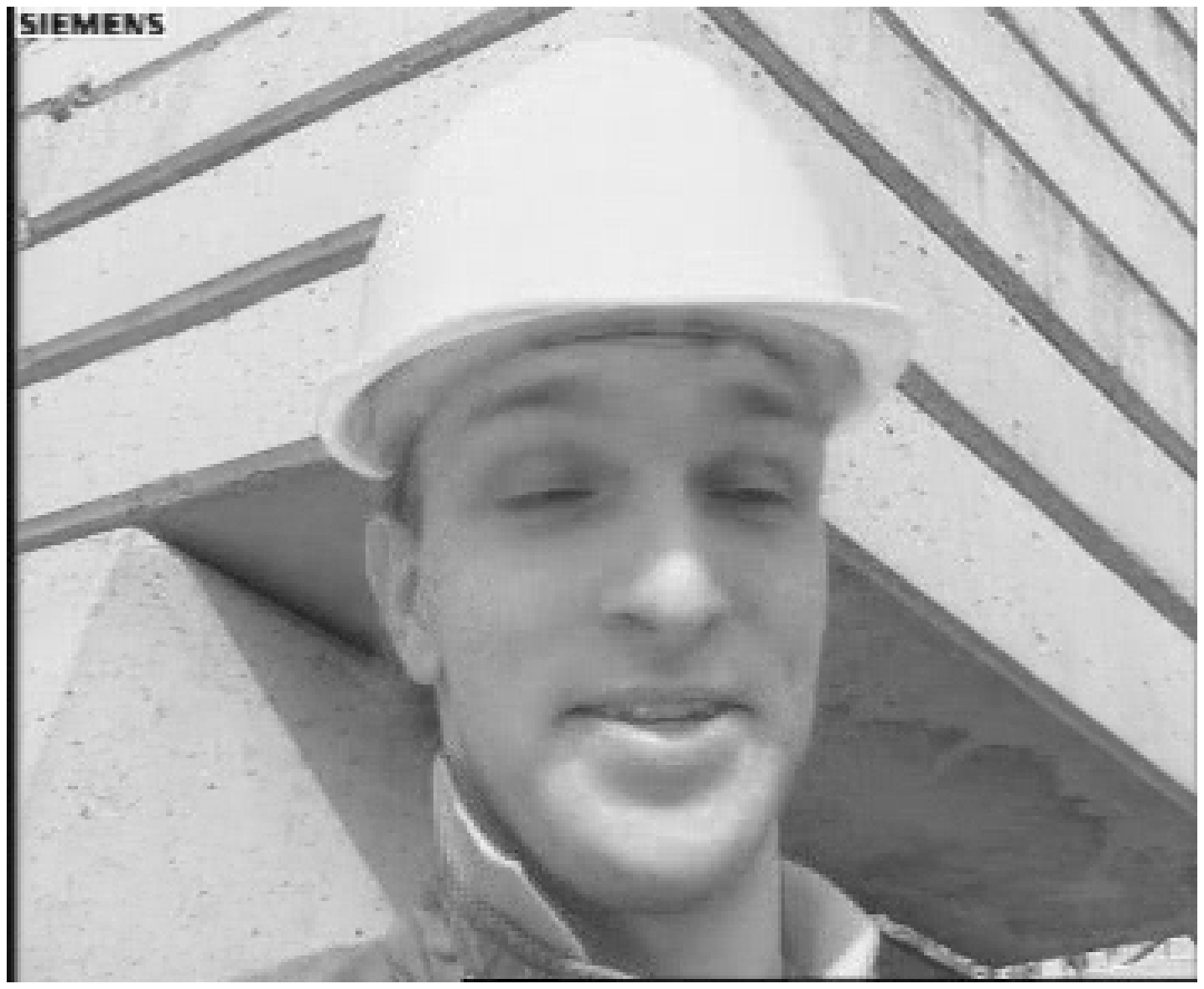}
        \caption{3D~MRDCT (PSNR=36.5764, MSSIM = 0.9280 )}
 \end{subfigure}
  \begin{subfigure}[t]{0.31\linewidth}
        \includegraphics[scale=0.35]{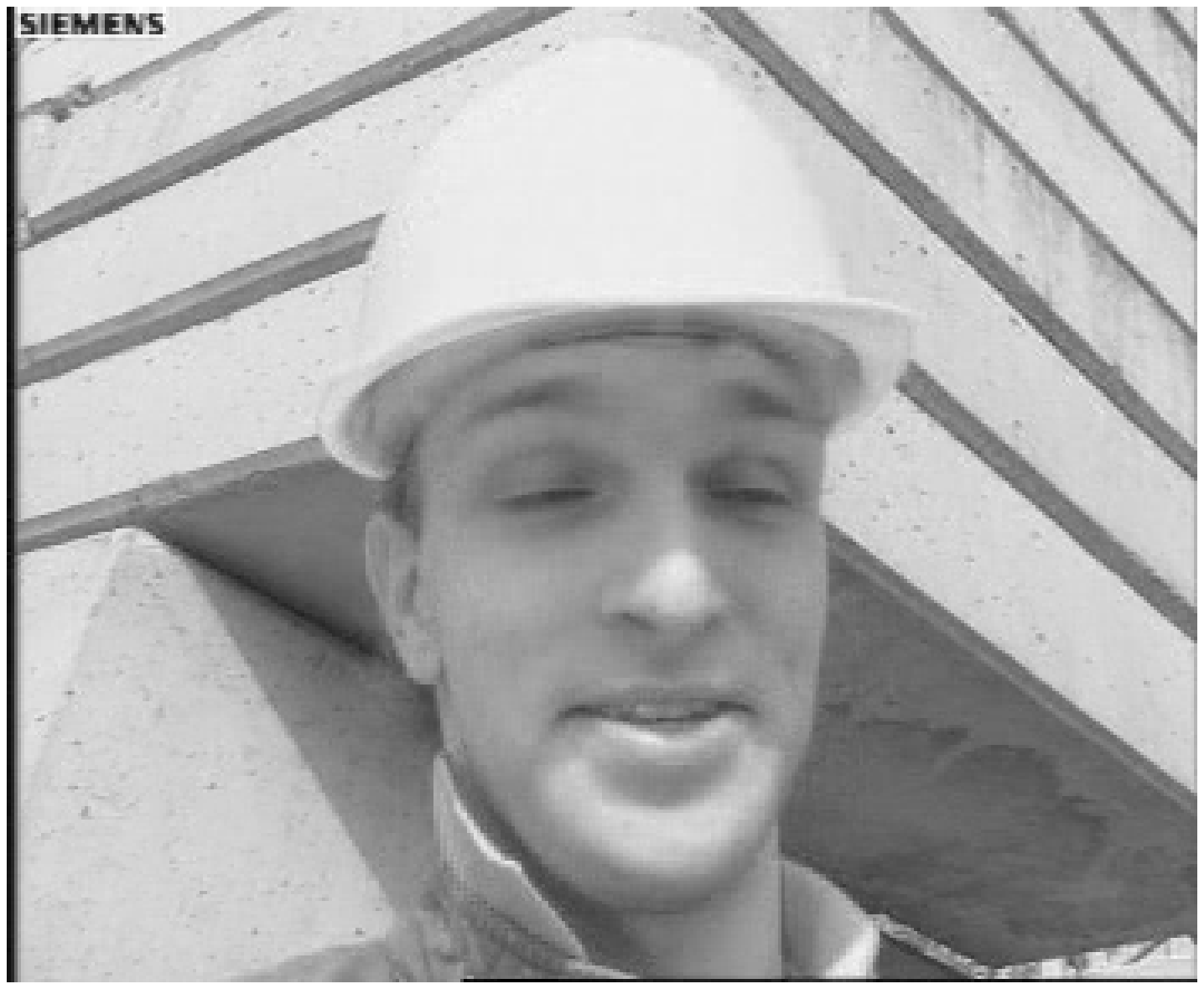}
        \caption{BAS-2008 (PSNR=37.0146, MSSIM = 0.9378)}
 \end{subfigure}
  \begin{subfigure}[t]{0.31\linewidth}
        \includegraphics[scale=0.35]{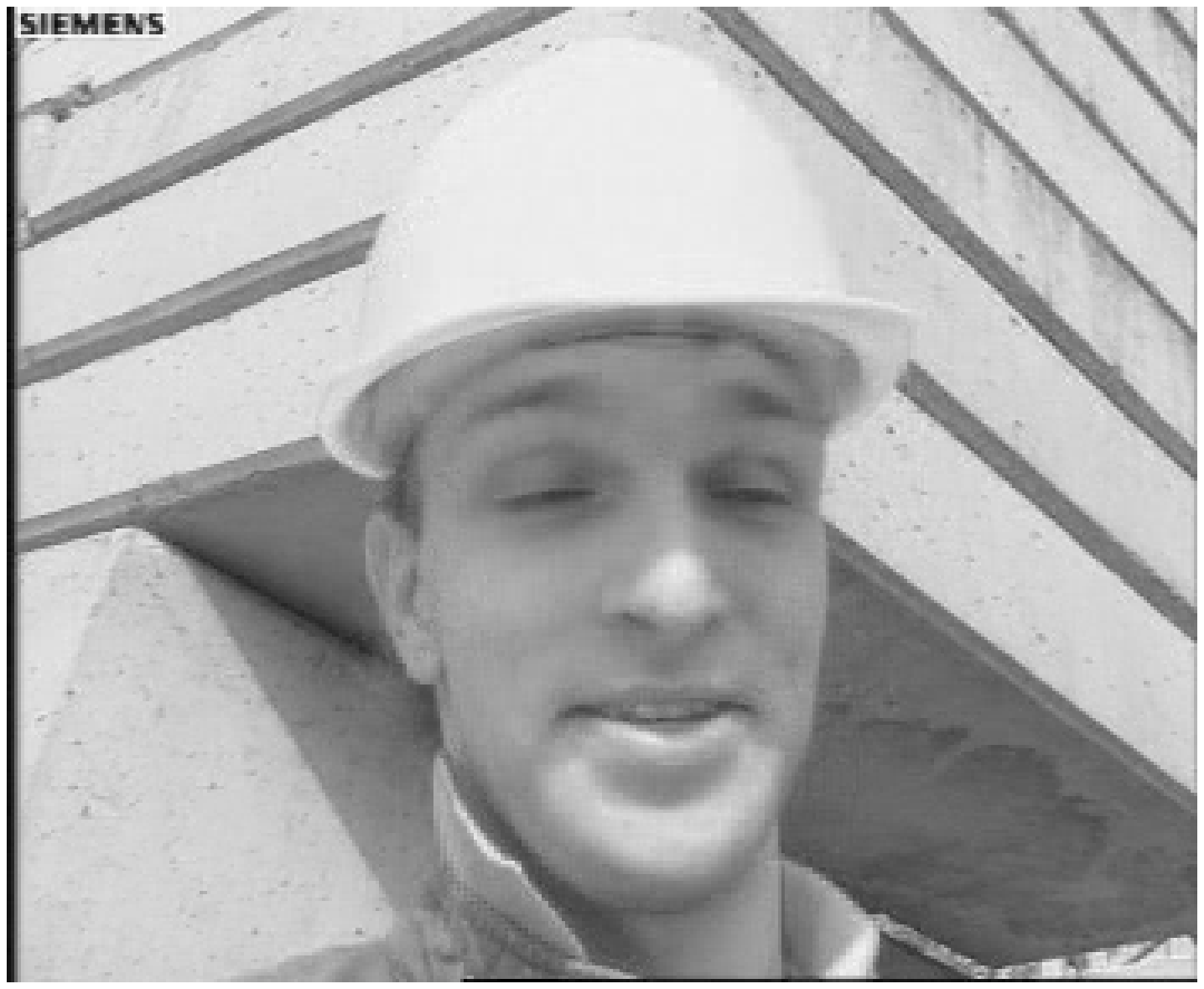}
        \caption{3D~BAS-2009 (PSNR=36.5367, MSSIM = 0.9341)}
 \end{subfigure}
  \begin{subfigure}[t]{0.31\linewidth}
        \includegraphics[scale=0.35]{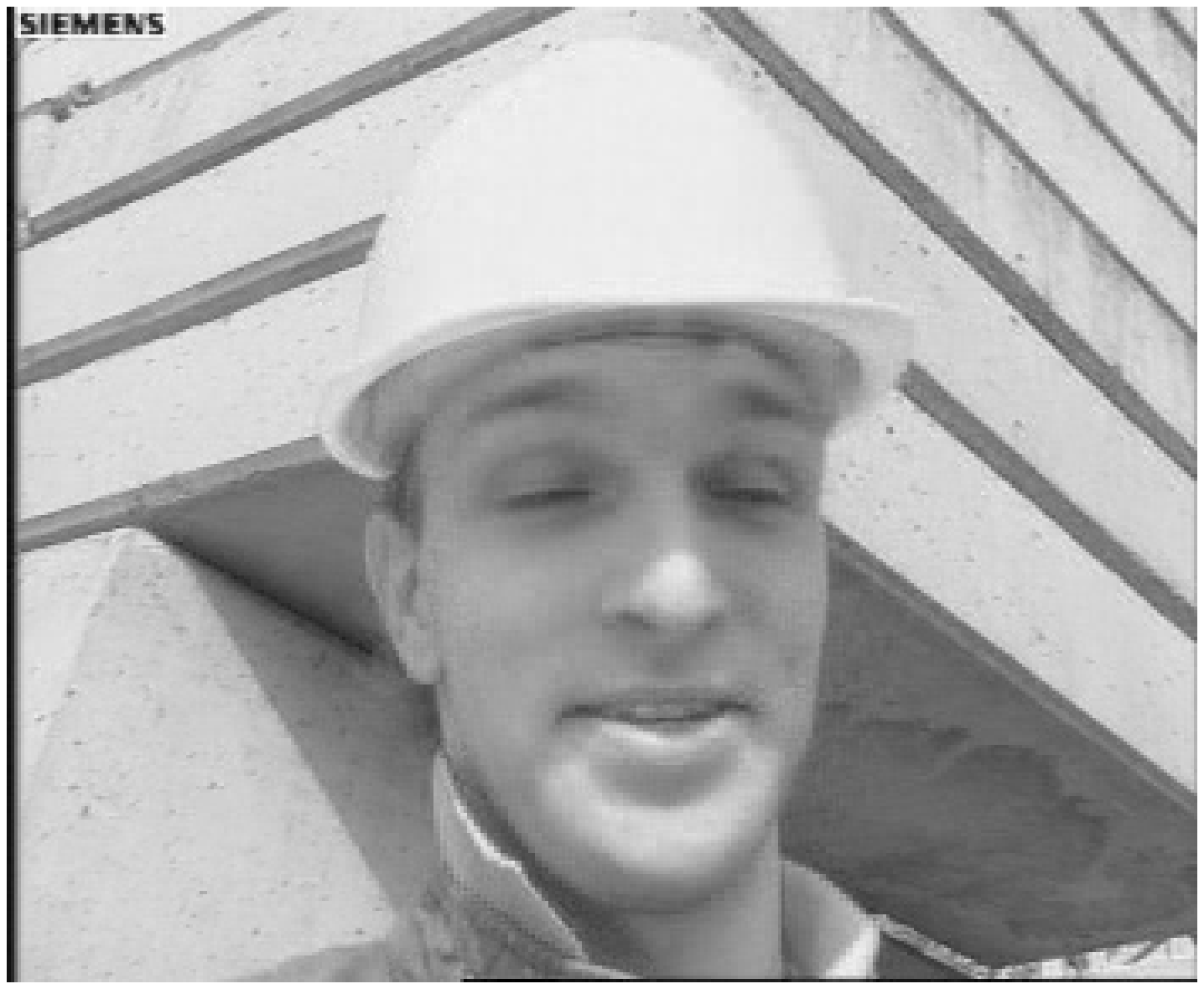}
        \caption{3D~BAS-2013 (PSNR=36.3304, MSSIM = 0.9323)}
 \end{subfigure}
  \begin{subfigure}[t]{0.31\linewidth}
        \includegraphics[scale=0.35]{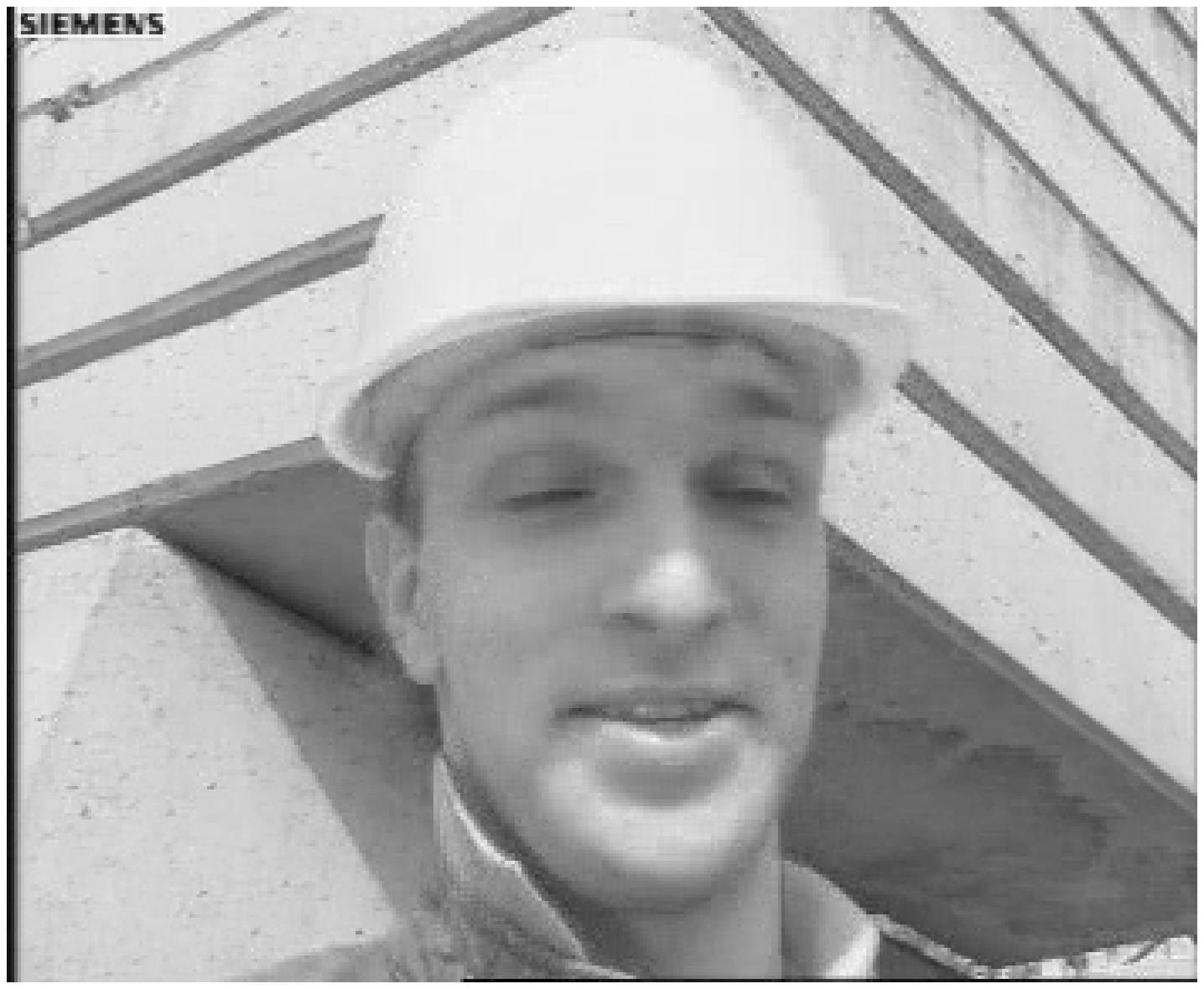}
        \caption{3D~IADCT (PSNR=35.2330, MSSIM = 0.9135)}
 \end{subfigure}
\caption{Qualitative assessment for the first frame of the ``foreman'' video sequence.
The complete compressed video sequence is available in~\cite{foreman_compressed}.}
\label{fig_qualitative}
\end{figure*}

The
average PSNR and
average MSSIM
for
all
considered
video sequences
are shown in \figurename~\ref{fig_perfomance}
at the discussed range of $Q$ values.
The 3D~DCT effected
the highest video quality,
whereas
the 3D~SDCT presented
a considerably lower performance
when compared to competing approximations.
Most approximate methods
performed similarly to the exact 3D~DCT,
specially at
low compression rate (small values of $Q$).
In the following analysis,
we exclude the 3D~SDCT
since
competing approximations showed
considerable higher performance.
Generally,
the approximate methods presented less than 7\% PSNR degradation
relative to the exact DCT.
For example,
for $Q=0.1$,
the 3D~LODCT shows only $1.53\%$ lower PSNR compared to the exact DCT,
whereas
the
3D~IADCT presents a $5.22\%$ PSNR reduction.
For $Q=4$,
such values are $2.01\%$ and $6.36\%$, respectively.
\figurename~\ref{fig_qualitative}
presents
a qualitative comparison
for the first frame of the ``foreman'' video sequence
for $Q=0.25$.
The frames are essentially indistinguishable.
The complete compressed video sequence is available in~\cite{foreman_compressed}.

We assessed
the
impact of
the motion level
on the video encoding performance
for each discussed methods
and
for all considered video sequences.
As the motion level measure,
we adopted the average motion vectors magnitude
$ \overline{  \lvert \mathbf{m} \rvert  }$~\cite{dawood1999content}.
The motion vectors were extracted from video frames
employing the ARPS~motion estimation algorithm~\cite{nie2002adaptive,barjatya2004block}.
In Table~\ref{tab_motion_level},
we show
the motion level of
each considered video sequence.

It is expected that the 3D~DCT
presents a performance decrease in higher motion level videos
since high temporal variation leads to
energy dispersion in higher 3D~DCT
coefficients~\cite{li2016spatiotemporal}.
However,
other factors such as the spatial texture of frames also impacts
on the visual quality~\cite{dawood1999content}.
Since 3D~DCT approximations
also aim at preserving
the original 3D~DCT properties,
such behavior is expected to happen in similar fashion.
We show in Table~\ref{tab_motion_level_perform}
the performance of each 3D~method
at each video sequence,
as well as the correlation coefficient~$\rho$
between
the performance metric and the motion level measure.
Indeed,
the correlation values show an inverse tendency between performance and motion level,
specially for the SSIM metrics,
which better captures the perceived visual quality
when compared with the PSNR~\cite{Wang2004}.
Furthermore,
all approximate methods present similar correlation values to the
exact 3D~DCT.
Therefore,
the proposed 3D~methods
are suitable candidates
for replacing the exact 3D~DCT
in video coding at different motion levels.

\begin{table}[h]
\centering
\caption{Motion level of considered video sequences}
\label{tab_motion_level}
\begin{tabular}{c c c }
\toprule
Sequence index & Sequence & Motion level $\overline{ \lvert \mathbf{m} \rvert }$ \\
\midrule
1 & akyio 	& 0.0387 \\
2 & container &	0.0954 \\
3 & news	 &	0.2561 \\
4 & silent & 	0.4877 \\
5 & mobile &	 0.748 \\
6 & mother-daughter & 0.7594 \\
7 & hall-monitor	 &	0.805 \\
8 & coastguard &	 1.7614 \\
9 & foreman	& 2.345 \\
\bottomrule
\end{tabular}
\end{table}

\begin{table*}[t]
\centering
\caption{Performance of proposed methods for each video sequence and correlation with motion level}
\label{tab_motion_level_perform}
\begin{tabular}{l c  c c c c c c c c c c}
\toprule
\multirow{3}{*}{Measure} & \multirow{3}{*}{Method} & \multicolumn{9}{c}{Sequence index} & \multirow{3}{*}{correlation $\rho$}  \\
\cmidrule{3-11}
&  & 1 & 2 & 3 & 4 & 5 & 6 & 7 & 8 & 9 &  \\
\midrule
\multirow{9}{*}{PSNR~(dB)} & 3D~DCT & 39.12 & 35.87 & 36.33 & 34.93 & 26.42
& 37.74 & 35.68 & 30.22 & 31.69
& -0.55 \\
& 3D~SDCT & 34.31 & 31.41 & 31.31 & 30.74 &
21.05 & 33.27 & 32.16 & 25.77 & 27.08 & -0.51  \\
& 3D~LODCT & 38.27 & 35.21 & 35.59 & 34.42 &
25.86 & 37.07 & 35.24 & 29.79 & 31.18 & -0.54  \\
& 3D~RDCT & 37.80 & 34.83 &  35.06 & 34.06 &
25.11 & 36.66 & 34.95 & 29.36 & 30.82 & -0.52 \\
& 3D~MRDCT & 36.71 & 33.68 & 33.88 & 33.13 & 23.42 &
35.58 & 34.28 & 28.13 & 29.61 & -0.51  \\
& 3D~BAS-2008 & 37.88 & 34.76 & 35.22 & 34.12 & 24.94 & 36.68
& 35.06 & 29.24 & 30.75 & -0.53 \\
& 3D~BAS-2009 & 37.52 & 34.50 & 34.83 & 33.87 & 24.50 & 36.34
& 34.81 & 28.96 & 30.44 & -0.52 \\
& 3D~BAS-2013 & 37.69 & 34.73 & 35.03 & 34.10 & 25.14 &
36.50 & 34.90 & 29.29 & 30.63 & -0.54 \\
& 3D~IADCT & 36.56 & 33.53 & 33.76 & 32.99 & 23.26 & 35.41 &
34.17 & 28.02 & 29.51 & -0.50 \\
\midrule
\multirow{9}{*}{SSIM} & 3D~DCT &
0.96 & 0.93 & 0.95 & 0.92 & 0.87 & 0.94 & 0.93 & 0.86 & 0.86 & -0.84 \\
& 3D~SDCT & 0.92 & 0.89 & 0.90 & 0.84 & 0.72 & 0.87 & 0.89 & 0.74 & 0.73 & -0.78 \\
& 3D~LODCT & 0.95 & 0.93 & 0.94 & 0.91 & 0.86 & 0.93 & 0.93 & 0.85 & 0.84 & -0.84  \\
& 3D~RDCT & 0.95 & 0.92 & 0.94 & 0.91 & 0.84 & 0.92 & 0.92 & 0.84 & 0.84 & -0.82 \\
& 3D~MRDCT & 0.94 & 0.91 & 0.93 & 0.89 & 0.79 & 0.91 & 0.92 & 0.80 & 0.80 & -0.76 \\
& 3D~BAS-2008 & 0.95 & 0.92 & 0.94 & 0.91 & 0.84 & 0.93 & 0.93 & 0.83 & 0.84 & -0.80 \\
& 3D~BAS-2009 & 0.95 & 0.92 & 0.94 & 0.91 & 0.83 &  0.92 & 0.92 & 0.82 & 0.83 & -0.80  \\
& 3D~BAS-2013 & 0.95 & 0.92 & 0.94 & 0.91 & 0.84 & 0.92 & 0.92 & 0.83 & 0.83 & -0.84 \\
& 3D~IADCT & 0.94 & 0.91 & 0.93 & 0.89 & 0.78 & 0.91 & 0.92 & 0.79 & 0.80 & -0.76 \\
\bottomrule
\end{tabular}
\end{table*}

\color{black}

\section{Visual Tracking}
\label{sec:tracking}

\begin{figure*}[t]
\centering
 \begin{subfigure}[t]{0.42\linewidth}
        \includegraphics[scale=0.33]{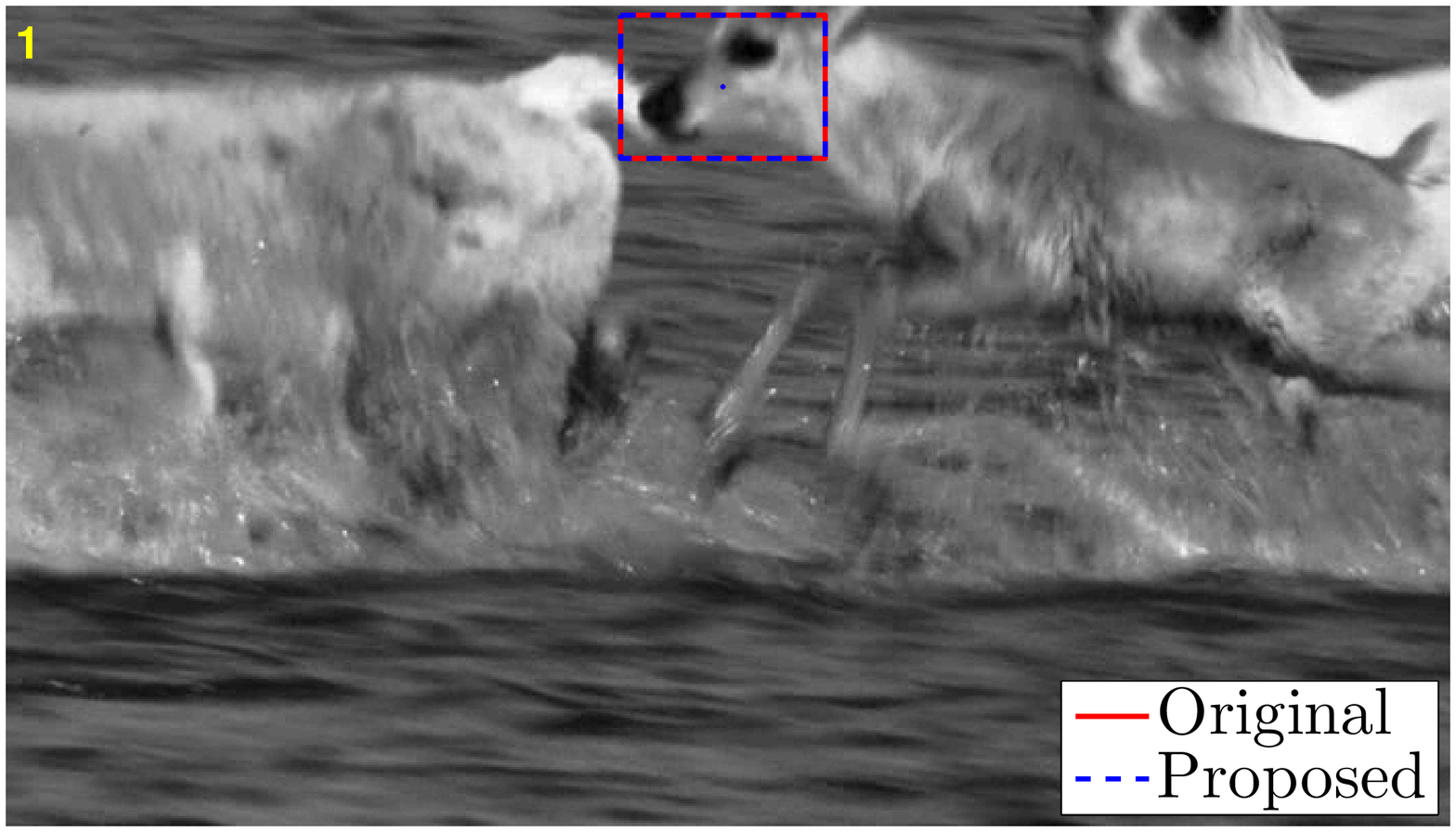}
 \end{subfigure}
 \begin{subfigure}[t]{0.42\linewidth}
        \includegraphics[scale=0.33]{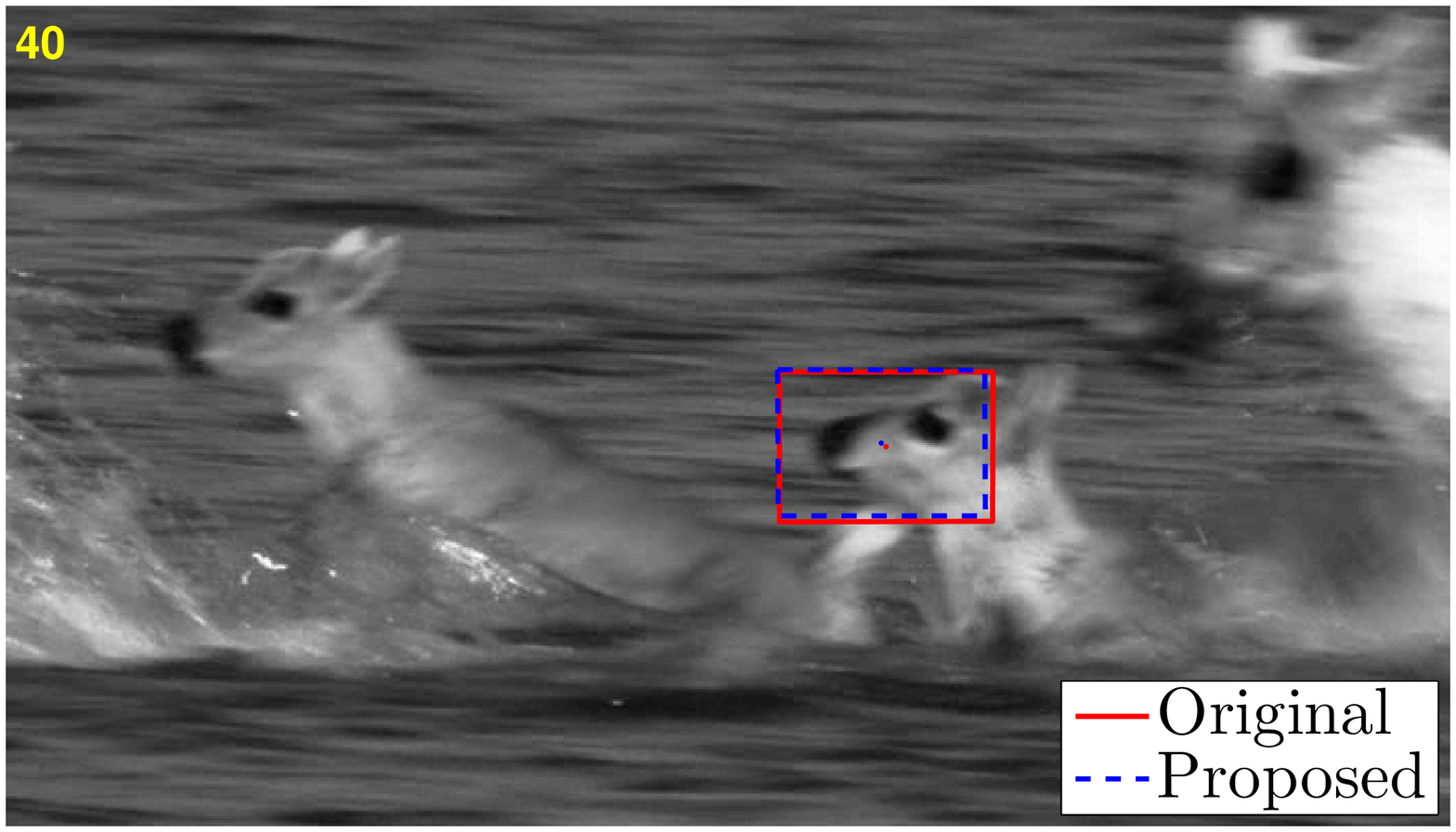}
 \end{subfigure}
 \\
  \begin{subfigure}[t]{0.42\linewidth}
        \includegraphics[scale=0.33]{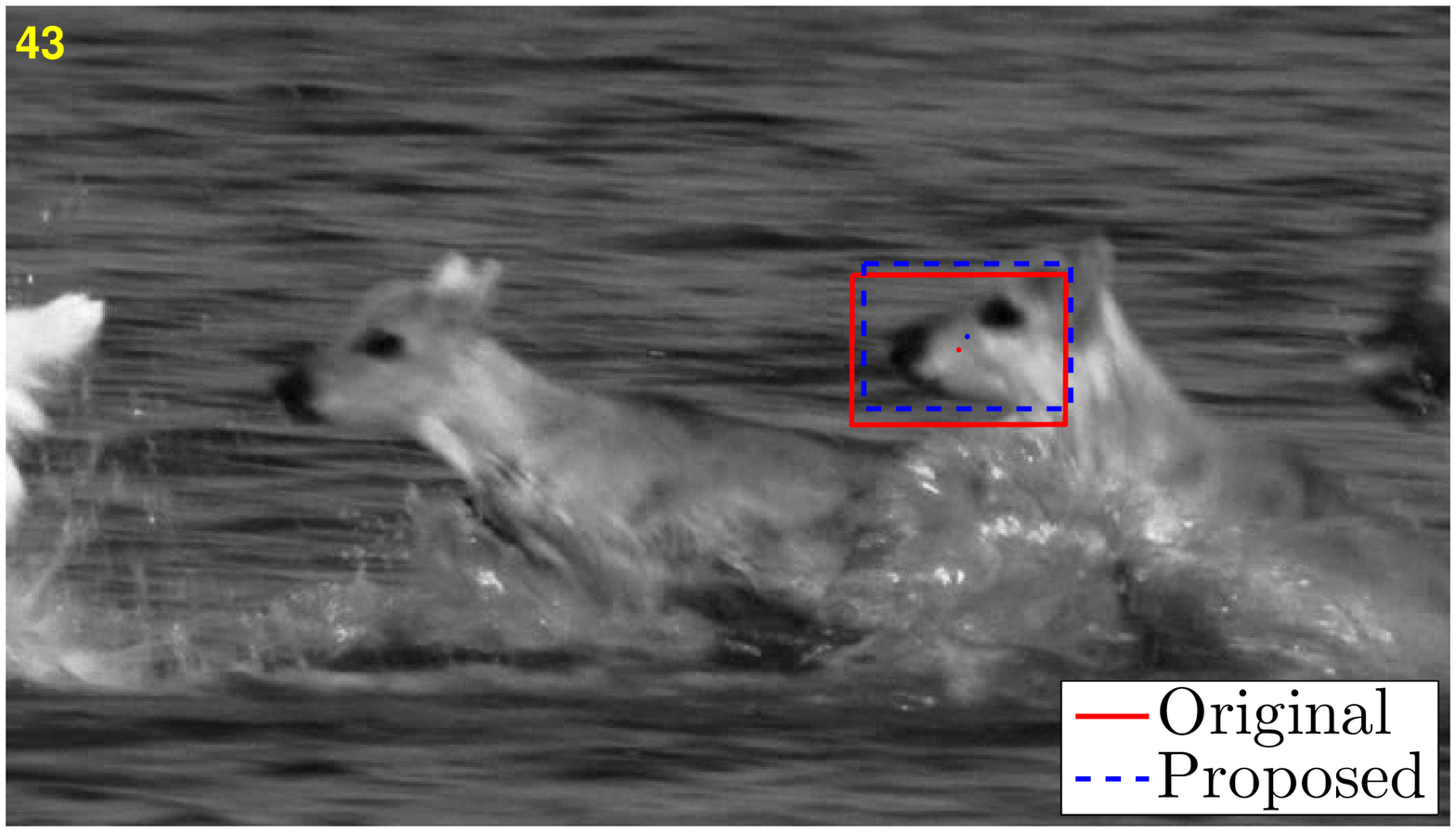}
 \end{subfigure}
  \begin{subfigure}[t]{0.42\linewidth}
        \includegraphics[scale=0.33]{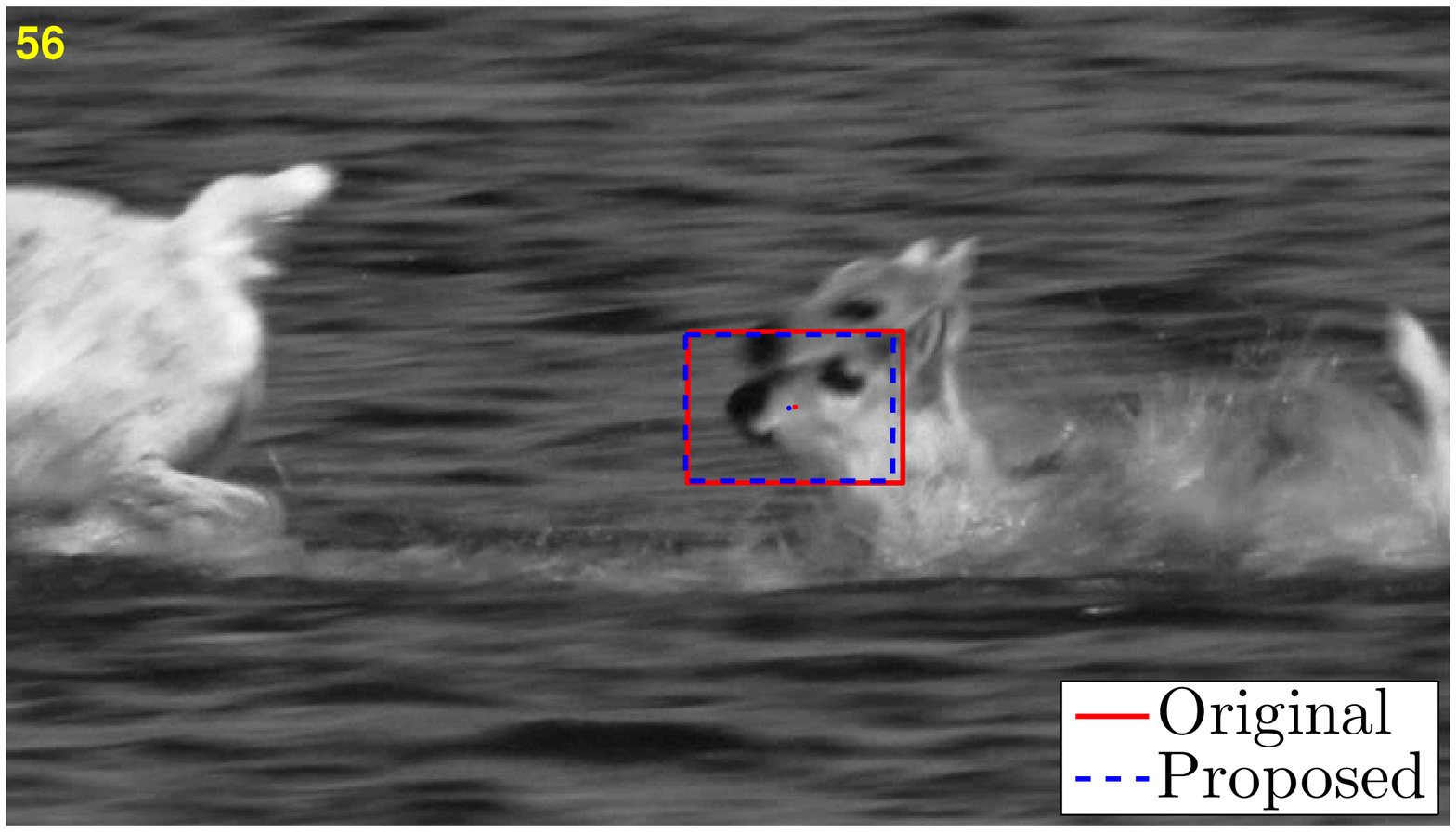}
 \end{subfigure}
 \\
  \begin{subfigure}[t]{0.42\linewidth}
        \includegraphics[scale=0.33]{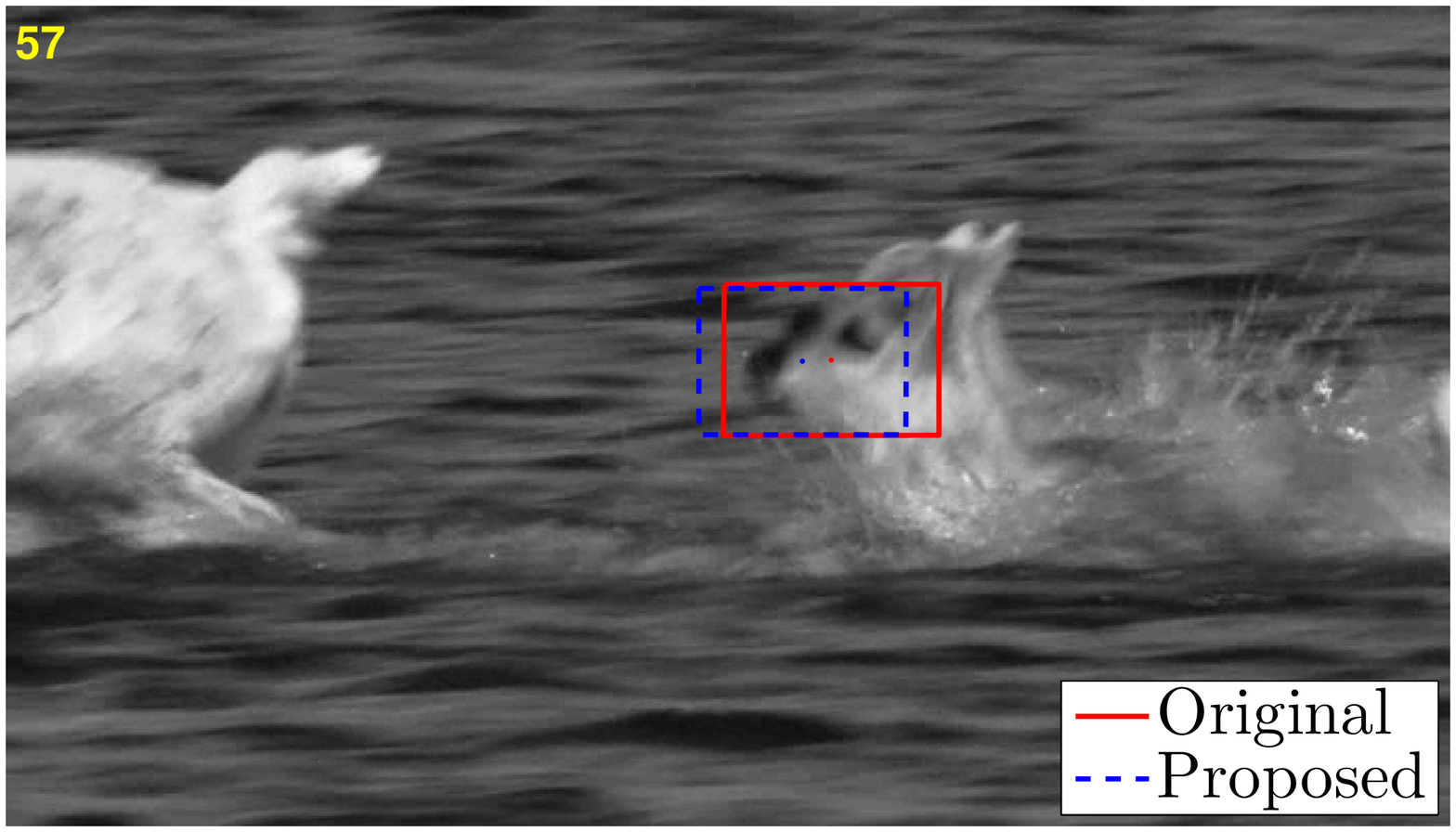}
 \end{subfigure}
  \begin{subfigure}[t]{0.42\linewidth}
        \includegraphics[scale=0.33]{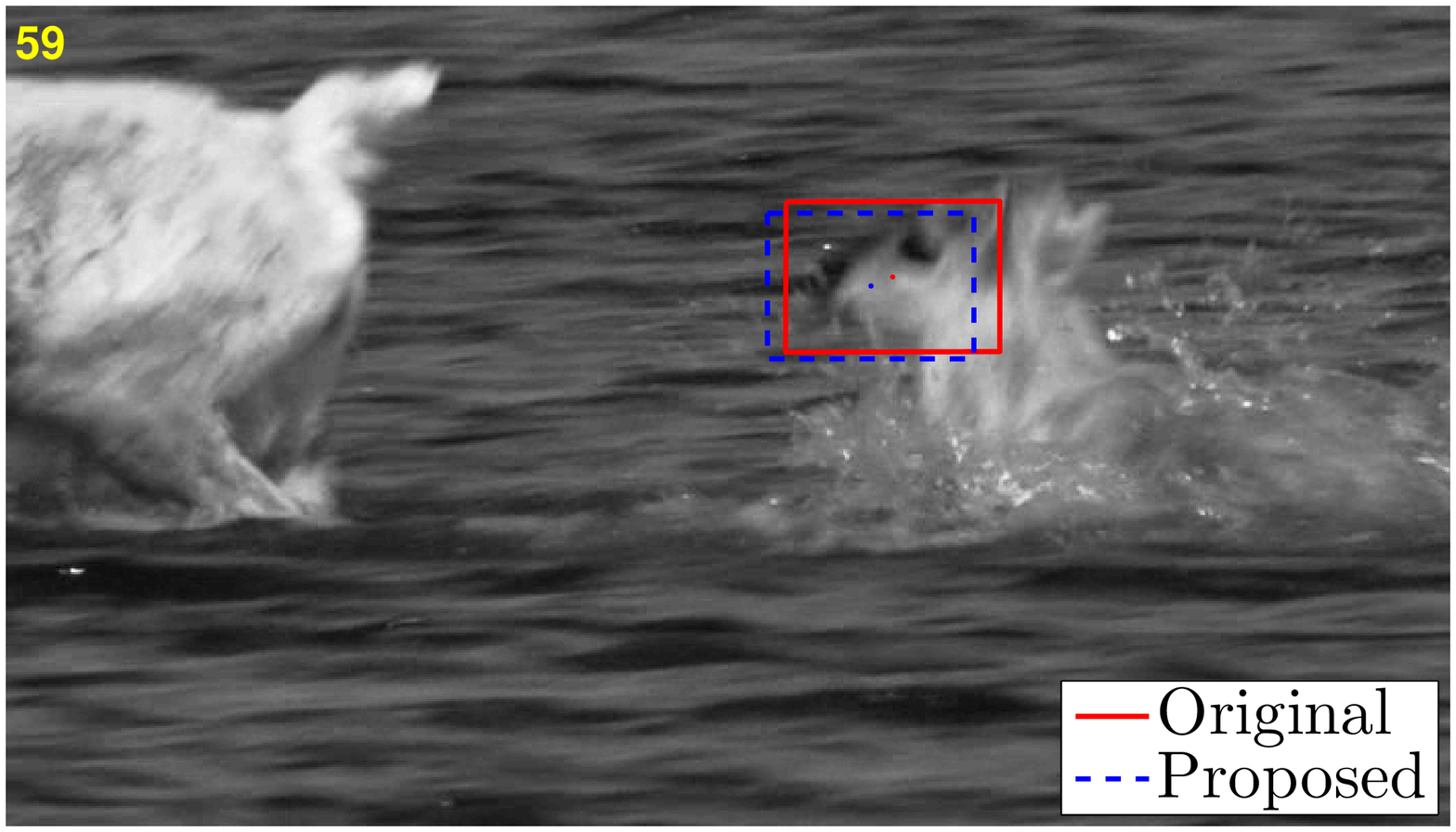}
 \end{subfigure}
 \caption{Qualitative tracking results for the ``animal'' video sequence for representative frames (first, 40th, 43rd, 56th, 57th, and 59th).
The full video sequence is available in~\cite{animal_tracked}.}
\label{fig_tracking_qualit}
\end{figure*}

\begin{figure*}[t]
\centering
        \includegraphics[scale=1]{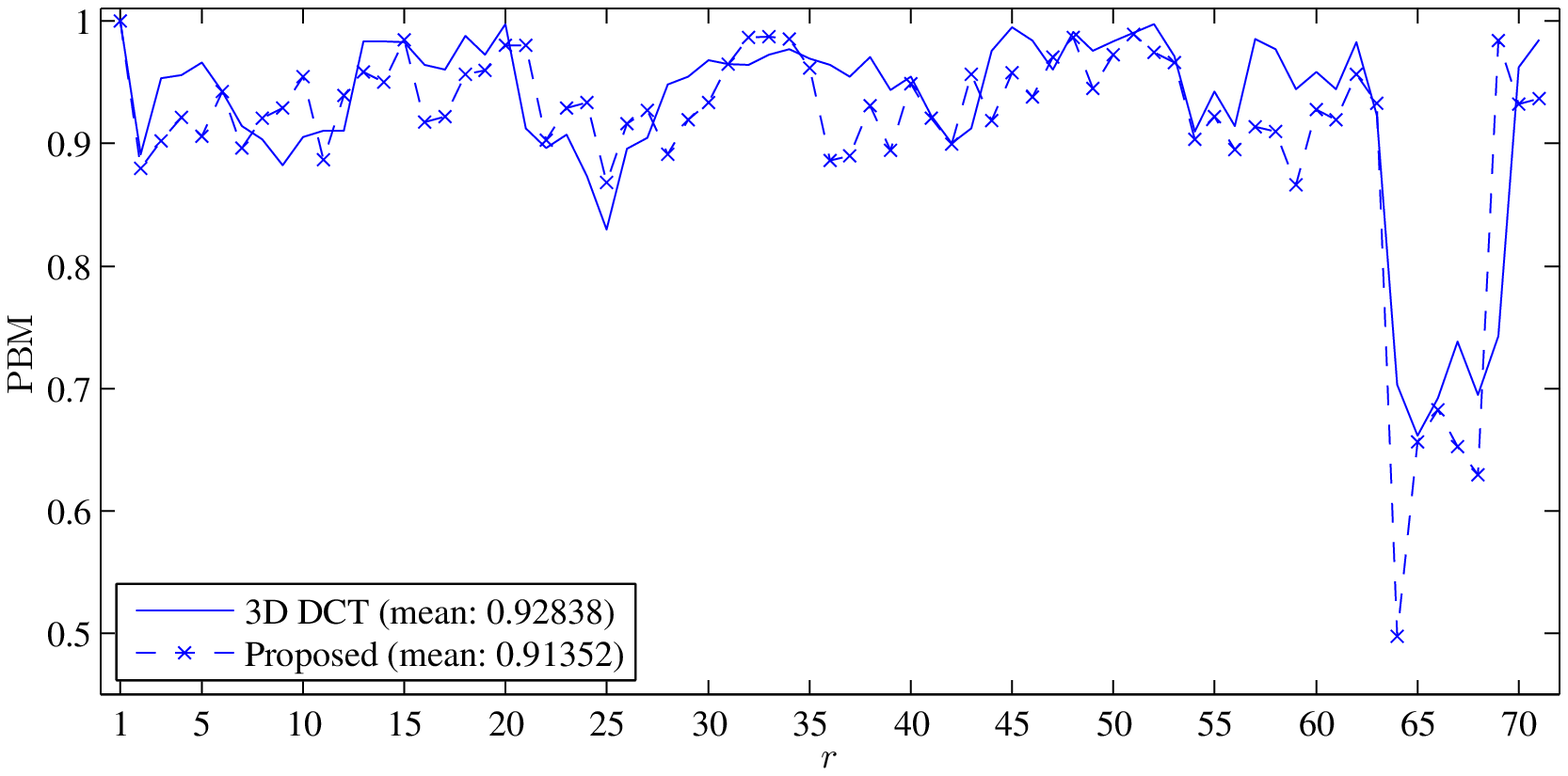}
        \caption{PBM for each $r$th frame of ``animal'' video sequence.}
\label{fig_tracking_quant}
\end{figure*}

Visual tracking consists in
predicting the location
of a target over a frame sequence
based on an initial target position.
A large variety of tracking methods
is available in literature~\cite{smeulders2014survey}.
Immediate applications include:
visual surveillance and security control~\cite{coifman1998tracking,javed2002tracking,wang2009distributed},
driver assistance and tracking vehicles~\cite{coifman1998tracking,almagambetov2015robust},
and
wireless sensor visual networks~(WSVN)~\cite{wang2009distributed,demigha2013tracking}.
The class of trackers
discussed in~\cite{smeulders2014survey,ross2008incremental,abdi2010principal}
employs principal component analysis~(PCA) as appearance model~\cite{smeulders2014survey,ross2008incremental}.
Mathematically,
PCA is equivalent to the KLT~\cite{gerbrands1981relationships}.
Inspired by PCA-based trackers and
by the relation between the KLT and the DCT,
Li~\emph{et al.} proposed
a discriminative learning-based
visual tracking method
based on the 3D~DCT
for low-dimensional subspace representation~\cite{li2013visualtracking}.
Such algorithm leads to
a computationally
more efficient implementation
when compared to
purely PCA-based techinques~\cite{li2013visualtracking, ross2008incremental}.

Low-complexity algorithms for visual tracking systems
paves the way
for real-time computationally demanding applications~\cite{chang2012real,almagambetov2015robust,zhou2015robust}.
As extensively discussed in~\cite{Liang2001,tran2000bindct,britanak2007discrete,haweel2001,lengwehasatit2004scalable,cb2011,bc2012,bas2008,bas2009,bas2013, Potluri2013,cintra2011integer,Cintra2014-sigpro,coutinho2015multiplierless,cintra2015energy,hnativ2014integer, reznik2007low},
DCT approximations are emerging tools for
DCT-based technologies
at
a
low computational cost.
In this section,
we embedded a 3D~DCT approximation
in a video tracking system.
We modified the 3D~DCT block in~\cite{li2013visualtracking}
in order to compute a 3D~DCT approximation,
according to~\eqref{3d_dct_approx2}.
The original algorithm in~\cite{li2013visualtracking}
computes $N_1 \times N_2 \times N_3$ incremental 3D~DCT
by using a fast Fourier transform~(FFT)~\cite{Ahmed1974}.
Quantities
$N_1$ and $N_2$
were set fixed,
whereas
$N_3$ varied incrementally
until a maximum buffer size~$T$.
The method demands
$N_1  N_2 \left( \log_2 N_1 + \log_2 N_2 \right) +N_1 N_2 N_3 \log_2 N_3$
complex multiplications
and twice such value of complex additions~\cite{li2013visualtracking,Blahut2010}.
We set $N_1 = N_2 = T = 8$ in order
to achieve compatibility
with the approximate methods considered
in the current work.

Among the DCT approximations shown in
Table~\ref{tab_dct_approx},
we selected the MRDCT~\cite{bc2012},
which possesses the lower additive complexity
as shown in Table~\ref{table_complexity}.
The IADCT~\cite{Potluri2013} presents the same computational cost,
but shows less favorable performance values in 3D~video compression simulations.
For the transient first frames,
where $N_3 < T$,
we employed a combined approximate/exact DCT algorithm
using~\eqref{multd_dct_approx} with $R=3$.
We utilized the MRDCT matrix for
$\hat{\mathbf{C}}_{N_1}$
and
$\hat{\mathbf{C}}_{N_2}$
and exact DCT matrix ${\mathbf{C}}_{N_3}$
instead of $\hat{\mathbf{C}}_{N_3}$
until $N_3$ reaches
the final value $N_3 = T = 8$.
Then,
we computed the 3D~MRDCT approximation
as proposed in~\eqref{3d_dct_approx1} and~\eqref{3d_dct_approx2}
for all the remaining frames.
In this case,
the original tracker demands
1920 and 3840
complex multiplications and complex additions,
respectively,
whereas the modified tracker
requires only
2688 real additions,
as shown in
Table~\ref{table_complexity}.
All remaining
video tracking
parameters were preserved
for both original and modified methods.

It is worth mentioning that
our main objective is
to
provide a proof-of-concept
for the proposed methods;
suggesting
low-complexity approximations
as a feasible
approach for video tracking computation.
Consequently,
we selected a representative
3D~DCT based video tracking system~\cite{li2013visualtracking}
for analysis.
\figurename~\ref{fig_tracking_qualit}
shows a qualitative
comparison for some
representative frames of
the ``animal'' video sequence,
from data set available in~\cite{tracking_database}.
Both original and proposed trackers
show very close performance.
The full video sequence is available in~\cite{animal_tracked}.

For a quantitative evaluation,
the
\emph{position-based measure}~(PBM)~\cite{smeulders2014survey} was computed.
This measure
is based on the distance of
the tracked bounding box centroid
relative to
a previously defined \emph{ground truth} bounding box centroid.
The employed ground truth data are provided in~\cite{tracking_database}.
Let
$\mathbb{T}_r$
and
$\mathbb{G}_r$
be
the
tracked
and
ground truth bounding boxes,
respectively,
for the $r$th video frame.
The PBM is given by~\cite{smeulders2014survey}:
\begin{align}
\operatorname{PBM}(r)
=
1-
\frac{\operatorname{D}(r)}
{T_h(r)}
,
\end{align}
where
$T_h(r) = [
\operatorname{width}
\left( \mathbb{T}_r \right)$
$+$
$
\operatorname{height}
\left( \mathbb{T}_r \right)$
$+$
$
\operatorname{width}
\left( \mathbb{G}_r \right)$
$+$
$
\operatorname{height}
\left( \mathbb{G}_r \right)
 ]/2
$,
and
\begin{align}
\operatorname{D}(r)
=
\begin{cases}
\parallel
\operatorname{cent} \left(\mathbb{T}_r  \right)
-
\operatorname{cent} \left(\mathbb{G}_r  \right)
\parallel,
&
\text{if } \mathbb{G}_r \cap \mathbb{T}_r \neq \varnothing,
\\
T_h(r), & \text{otherwise;}
\end{cases}
\end{align}
and
$\operatorname{cent}(\cdot)$,
$\operatorname{width}(\cdot)$,
and
$\operatorname{height}(\cdot)$
return
the
centroid, width, and height
of the bounding box argument,
respectively.
The PBM values are confined to the interval $[0,1]$.
Value~0 represents a tracking failure and
value~1 indicates that the centroid of $\mathbb{T}_r$ and $\mathbb{G}_r$
are the same.
Values close to 1 indicates good tracking performance.
\figurename~\ref{fig_tracking_quant}
shows the PBM for all frames of
the ``animal'' video sequence.
The proposed method values tend to follow the original method curve,
being sometimes even higher.
In average,
the proposed method
present only $1.6\%$ lower PBM value
at a much lower complexity cost,
as shown in
Table~\ref{table_complexity}
and
Table~\ref{table_complexity_reduction}.

\section{Conclusion}
\label{sec:conclusion}

In the current work,
an algebraic formulation for
linear 3D~DCT approximations
was proposed
in terms of tensor analysis.
Several multiplierless 3D~DCT approximations were suggested
based on state-of-art approximate DCT matrices.
The concept was generalized
and multidimensional approximate DCT
were formulated.
Mathematical expressions for the multidimensional complexity cost were
derived.
Such expressions were considered
in the 3D case to
assess the
computational overhead
of the proposed methods.

The approximations were applied
for interframe video coding.
In such context,
we proposed
a procedure to modify the quantization volume
in order to facilitate
the use of low-complexity 3D~DCT approximations.
The obtained results
showed
that 3D~DCT approximations present
competitive performance compared to exact 3D~DCT
algorithms
at a considerably lower computational cost.
We also simulated a video tracking system,
originally based on the exact 3D~DCT,
embedded with 3D~DCT approximations.
Results showed
that the modified low-cost method
performed very closely
to the original method.
We conclude that
3D~DCT approximations
can be effective low-complexity tools
for emerging hardware and energy limited 3D~DCT-based technologies,
and also
for
current systems
that benefit of 3D~DCT computation,
such as SoftCast broadcasting system~\cite{softcast2010,jakubczak2010softcast,jakubczak2011softcast},
3D~quantization-based video encoders~\cite{mulla2014image,Saponara2012,
sawant2011balanced,bozinovic2003scan,lai2002video,chan1997variable, lee1997quantization},
transform-based visual trackers~\cite{li2013visualtracking}.
\color{black}

\appendix[Proof of Equation~\eqref{tensor_y}]

Let us split~\eqref{3d_dct_approx2}
into two parts:
(i)~the $i$-mode products involving only low-complexity matrix $\mathbf{T}_N$
and
(ii)~the $i$-mode products requiring the diagonal matrix~$\mathbf{S}_N$.
Part~(i)
is given in~\eqref{tensor_A}.
The next intermediate tensor is
\begin{align}
\mathcal{B} =
\mathcal{A}
\times_1
\mathbf{S}_N.
\label{tensor_B}
\end{align}
Let the diagonal matrix $\mathbf{S}_N$
entries be:
\begin{align}
\label{diagonal_elements}
s[k,n]
\triangleq
\begin{cases}
d_k, \quad \text{if } k = n, \\
0, \quad \text{otherwise},
\end{cases}
\quad
n,k = 0,1,\ldots,N-1,
\end{align}
where $d_k$ is the $k$th diagonal element of matrix $\mathbf{S}_N$.
Taking into account~\eqref{general_tensor_prod},
tensor $\mathcal{B}$ entries are given by:
\begin{align}
b[k_1, n_2, n_3] =
\sum_{n_1 = 0}^{N-1}
a[n_1,n_2,n_3]
\cdot
s[k_1,n_1] .
\label{tensor_b_1}
\end{align}
Replacing~\eqref{diagonal_elements} into~\eqref{tensor_b_1},
we derive:
\begin{align}
b[k_1, n_2, n_3] =
a[k_1, n_2, n_3]
\cdot
d_{k_1}
.
\label{tensor_b_2}
\end{align}
Now the next intermediate tensor
is
given by:
\begin{align}
\mathcal{C} =
\mathcal{B}
\times_2
\mathbf{S}_N.
\label{tensor_C}
\end{align}
In a similar manner,
$\mathcal{C}$ entries are furnished by:
\begin{align}
\begin{split}
c[k_1, k_2, n_3] &=
b[k_1, k_2, n_3]
\cdot
d_{k_2} \\
& = a[k_1, k_2, n_3]
\cdot
d_{k_1}
\cdot
d_{k_2}
.
\end{split}
\label{tensor_c}
\end{align}
Finally,
the tensor $\mathcal{Y}$ given in~\eqref{3d_dct_approx2}
is also expressed by:
\begin{align}
\mathcal{Y} =
\mathcal{C}
\times_3
\mathbf{S}_N,
\label{tensor_Y}
\end{align}
whose entries are analogously obtained by:
\begin{align}
\begin{split}
y[k_1, k_2, k_3] &=
c[k_1, k_2, k_3]
\cdot
d_{k_3} \\
& = a[k_1, k_2, k_3]
\cdot
d_{k_1}
\cdot
d_{k_2}
\cdot
d_{k_3}.
\end{split}
\end{align}

\section*{Acknowledgments}

This work was partially supported by
the
CPNq,
Brazil.

\onecolumn

{\small
\singlespacing
\bibliographystyle{siam}
\bibliography{ref}
}

\end{document}